\documentclass[12pt]{article}
\usepackage{amssymb}
\usepackage{graphicx}
\usepackage{xcolor}

\definecolor{primary}{RGB}{0, 139, 139}      
\definecolor{secondary}{RGB}{180, 40, 40}   
\definecolor{accent}{RGB}{40, 120, 120}      
\usepackage[colorlinks=true]{hyperref}
\hypersetup{linkcolor=primary,citecolor=accent,urlcolor=secondary}

\begin{document}
\newcommand{\beq}{\begin{equation}}
\newcommand{\eeq}{\end{equation}}
\newcommand{\beqa}{\begin{eqnarray}}
\newcommand{\eeqa}{\end{eqnarray}}
\newcommand{\beqar}{\begin{eqnarray*}}
\newcommand{\eeqar}{\end{eqnarray*}}
\newcommand{\al}{\alpha}
\newcommand{\be}{\beta}
\newcommand{\del}{\delta}
\newcommand{\D}{\Delta}
\newcommand{\eps}{\epsilon}
\newcommand{\ga}{\gamma}
\newcommand{\Ga}{\Gamma}
\newcommand{\ka}{\kappa}
\newcommand{\nn}{\nonumber}
\newcommand{\inn}{\!\cdot\!}
\newcommand{\h}{\eta}
\newcommand{\ii}{\iota}
\newcommand{\kk}{\varphi}
\newcommand\F{{}_3F_2}
\newcommand{\la}{\lambda}
\newcommand{\La}{\Lambda}
\newcommand{\na}{\prt}
\newcommand{\Om}{\Omega}
\newcommand{\om}{\omega}
\newcommand{\p}{\Phi}
\newcommand{\sig}{\sigma}
\renewcommand{\t}{\theta}
\newcommand{\z}{\zeta}
\newcommand{\ssc}{\scriptscriptstyle}
\newcommand{\eg}{{\it e.g.,}\ }
\newcommand{\ie}{{\it i.e.,}\ }
\newcommand{\labell}[1]{\label{#1}} 
\newcommand{\reef}[1]{(\ref{#1})}
\newcommand\prt{\partial}
\newcommand\veps{\varepsilon}
\newcommand{\pol}{\varepsilon}
\newcommand\vp{\varphi}
\newcommand\ls{\ell_s}
\newcommand\cF{{\cal F}}
\newcommand\cA{{\cal A}}
\newcommand\cS{{\cal S}}
\newcommand\cT{{\cal T}}
\newcommand\cV{{\cal V}}
\newcommand\cL{{\cal L}}
\newcommand\cM{{\cal M}}
\newcommand\cN{{\cal N}}
\newcommand\cG{{\cal G}}
\newcommand\cK{{\cal K}}
\newcommand\cH{{\cal H}}
\newcommand\cI{{\cal I}}
\newcommand\cJ{{\cal J}}
\newcommand\cl{{\iota}}
\newcommand\cP{{\cal P}}
\newcommand\cQ{{\cal Q}}
\newcommand\cg{{\tilde {{\cal G}}}}
\newcommand\cR{{\cal R}}
\newcommand\cB{{\cal B}}
\newcommand\cO{{\cal O}}
\newcommand\tcO{{\tilde {{\cal O}}}}
\newcommand\bz{\bar{z}}
\newcommand\bb{\bar{b}}
\newcommand\ba{\bar{a}}
\newcommand\bg{\bar{g}}
\newcommand\bc{\bar{c}}
\newcommand\bw{\bar{w}}
\newcommand\bX{\bar{X}}
\newcommand\bK{\bar{K}}
\newcommand\bA{\bar{A}}
\newcommand\bH{\bar{H}}
\newcommand\bF{\bar{F}}
\newcommand\bxi{\bar{\xi}}
\newcommand\bphi{\bar{\phi}}
\newcommand\bpsi{\bar{\psi}}
\newcommand\bprt{\bar{\prt}}
\newcommand\bet{\bar{\eta}}
\newcommand\btau{\bar{\tau}}
\newcommand\hF{\hat{F}}
\newcommand\hA{\hat{A}}
\newcommand\hT{\hat{T}}
\newcommand\htau{\hat{\tau}}
\newcommand\hD{\hat{D}}
\newcommand\hf{\hat{f}}
\newcommand\hK{\hat{K}}
\newcommand\hg{\hat{g}}
\newcommand\hp{\hat{\Phi}}
\newcommand\hi{\hat{i}}
\newcommand\ha{\hat{a}}
\newcommand\hb{\hat{b}}
\newcommand\hQ{\hat{Q}}
\newcommand\hP{\hat{\Phi}}
\newcommand\hS{\hat{S}}
\newcommand\hX{\hat{X}}
\newcommand\tL{\tilde{\cal L}}
\newcommand\hL{\hat{\cal L}}
\newcommand\tG{{\tilde G}}
\newcommand\tg{{\tilde g}}
\newcommand\tphi{{\widetilde \Phi}}
\newcommand\tPhi{{\widetilde \Phi}}
\newcommand\te{{\tilde e}}
\newcommand\tk{{\tilde k}}
\newcommand\tf{{\tilde f}}
\newcommand\tH{{\tilde H}}
\newcommand\ta{{\tilde a}}
\newcommand\tb{{\tilde b}}
\newcommand\tc{{\tilde c}}
\newcommand\td{{\tilde d}}
\newcommand\tm{{\tilde m}}
\newcommand\tmu{{\tilde \mu}}
\newcommand\tnu{{\tilde \nu}}
\newcommand\talpha{{\tilde \alpha}}
\newcommand\tbeta{{\tilde \beta}}
\newcommand\trho{{\tilde \rho}}
 \newcommand\tR{{\tilde R}}
\newcommand\teta{{\tilde \eta}}
\newcommand\tF{{\widetilde F}}
\newcommand\tK{{\tilde K}}
\newcommand\tE{{\widetilde E}}
\newcommand\tpsi{{\tilde \psi}}
\newcommand\tX{{\widetilde X}}
\newcommand\tD{{\widetilde D}}
\newcommand\tO{{\widetilde O}}
\newcommand\tS{{\tilde S}}
\newcommand\tB{{\tilde B}}
\newcommand\tA{{\widetilde A}}
\newcommand\tT{{\widetilde T}}
\newcommand\tC{{\widetilde C}}
\newcommand\tV{{\widetilde V}}
\newcommand\thF{{\widetilde {\hat {F}}}}
\newcommand\Tr{{\rm Tr}}
\newcommand\tr{{\rm tr}}
\newcommand\STr{{\rm STr}}
\newcommand\hR{\hat{R}}
\newcommand\M[2]{M^{#1}{}_{#2}}
\newcommand\MZ{\mathbb{Z}}
\newcommand\MR{\mathbb{R}}
\newcommand\bS{\textbf{ S}}
\newcommand\bI{\textbf{ I}}
\newcommand\bJ{\textbf{ J}}

\begin{titlepage}
\begin{center}

\vskip 0.5 cm
{\LARGE \bf 
Tachyon and   massless sectors of  the \\  \vskip 0.25 cm  bosonic string effective action  
} \\
\vskip 1.25 cm
 Mohammad R. Garousi \footnote{garousi@um.ac.ir}

\vskip 1 cm
{{\it Department of Physics, Faculty of Science, Ferdowsi University of Mashhad\\}{\it P.O. Box 1436, Mashhad, Iran}\\}
\vskip .1 cm
 \end{center}

\begin{abstract}

Recently, it has been observed that in certain schemes of the bosonic string effective action, the non-derivative tachyon appears only in the tachyon potential, while all other couplings involve covariant derivatives of the tachyon and massless fields. Within these schemes, a minimal basis can be found whose coupling constants are constrained by T-duality upon circle reduction. Using this constraint, one obtains sets of couplings at each order of \(\alpha'\) that are invariant under Buscher rules supplemented by higher-derivative corrections for all base-space fields except the tachyon. In this paper, we observe that consistency of these T-dual sets with S-matrix elements requires an appropriate field redefinition, such that each set becomes invariant under Buscher rules supplemented by higher-derivative corrections for all base-space fields except the metric, dilaton, and tachyon. This separates the couplings into a massless sector and a tachyon sector. We determine the six-derivative couplings in the massless sector, with all coupling constants fixed, and find two T-dual couplings in the tachyon sector with arbitrary coupling constants. We show that the leading-order action and the massless-sector couplings exactly reproduce the expansion of the sphere-level S-matrix element of two tachyons and two massless vertex operators at six-momentum order, and that the coupling constant of the four-tachyon coupling in the tachyon sector is fixed by the expansion of the four-tachyon S-matrix element at six-momentum order.

\end{abstract}

\end{titlepage}

\tableofcontents

\section{Introduction}

Perturbative bosonic string theory has been proposed as a candidate for quantum gravity because its spectrum contains a massless spin-2 particle, which is believed to be the graviton. However, since its spectrum in Minkowski spacetime also contains a scalar particle with imaginary mass—known as the tachyon—this indicates that Minkowski spacetime is an unstable vacuum. To circumvent this issue, one typically enhances the symmetry of the worldsheet or spacetime to supersymmetry, which extends the bosonic string to the superstring. In the superstring spectrum, the graviton is preserved, while supersymmetry removes the tachyon \cite{Becker:2007zj}. However, so far there is no observation indicating that nature employs any form of supersymmetry \cite{Constantin:2025mex}. Consequently, some researchers focus on non-supersymmetric string theory \cite{Baykara:2024tjr,Baykara:2026vdc,Baykara:2026gem,Altavista:2026evd,Altavista:2026brr}. In a recent paper, we found that the classical tachyon potential of bosonic string theory actually has a minimum, around which the bosonic string no longer contains a tachyon \cite{Garousi:2026imu}. This motivates us to study the effective action of bosonic string theory in greater detail.

The worldsheet theory possesses conformal symmetry, which allows one to compute correlation functions of various massless and tachyon states at the sphere level, torus level, and higher genus. From these, one can in principle determine an \(\alpha'\)-expansion or higher-covariant-derivative expansion for the spacetime effective action at each genus that reproduces the expansion of these correlation functions, or S-matrix elements, at various orders of the Mandelstam variables \cite{Scherk:1971xy,Scherk:1974ca,Gross:1986iv,Gross:1986mw}. Alternatively, conformal symmetry can be used to derive the equations of motion for the background fields in spacetime, from which one can obtain the above low-energy effective action \cite{Tseytlin:2000mt,Tseytlin:1991bu,Banks:1991sg}. More directly, when one spatial dimension is compactified on a circle, conformal symmetry causes the various components of the spacetime massless fields in the base space to transform into one another; these transformations are known as T-duality transformations, or Buscher rules \cite{Buscher:1987sk,Rocek:1991ps}. It turns out that the requirement that the diffeomorphism-invariant effective action of bosonic string theory also be invariant under these Buscher rules, supplemented by higher-derivative corrections, imposes strong constraints on the classical effective action, especially when one is interested only in the massless fields \cite{Garousi:2019wgz,Garousi:2019mca,Wulff:2024ips,Ameri:2025bei}. When both massless and tachyon fields are involved, one must use both the T-duality and the S-matrix methods to fully determine the effective action, because the tachyon is invariant under T-duality. Hence, one should use T-duality to determine the effective action up to certain parameters that T-duality cannot fix. Then one should use the S-matrix method to fix these remaining parameters \cite{Garousi:2026imu}.

To use T-duality, one first needs a minimal basis of couplings and then uses T-duality to fix the coupling constants of the basis \cite{Garousi:2019wgz}. It has been observed that consistency with string theory S-matrix elements requires the diffeomorphism-invariant effective action to exclude couplings with an odd number of tachyons \cite{Garousi:2026nvz}. It has also been observed that, because of the presence of the tachyon potential in the effective action, one can use an appropriate non-singular higher-derivative field redefinition for the tachyon field that absorbs all covariant-derivative couplings that also include the non-derivative tachyon. In these schemes, the non-derivative tachyon then appears only in the tachyon potential, while all other couplings involve covariant derivatives of the tachyon and massless fields \cite{Garousi:2026imu}. Within these schemes, one can then use arbitrary field redefinitions for the massless fields to find the minimal basis, and then use consistency of the circular reduction of the minimal basis with Buscher rules supplemented by arbitrary higher-derivative corrections to find the coupling constants in the basis \cite{Garousi:2026imu}. However, as we will see below, consistency with string theory S-matrix elements requires that the higher-derivative field redefinitions and higher-derivative corrections to the Buscher rules be restricted.

In studying the higher-derivative effective action, a higher-derivative field redefinition is equivalent to imposing the equations of motion into the  effective action. The tachyon equation of motion is \(\Delta T = 0\), where
\beqa
\Delta T&=&\nabla_\mu\nabla^\mu T-2\nabla_\mu\Phi\nabla^\mu T-2V'(T)+\cdots,\labell{Teom}
\eeqa
where the dots represent terms at higher-derivative order. A tachyon field redefinition means that multiplying \(\Delta T\) by all arbitrary contractions of covariant derivatives of massless and tachyon fields, \(\delta T\), is allowed to be added to the effective action. The presence of \(V'(T)\) in \reef{Teom} indicates that higher-derivative field redefinitions of the tachyon are no longer allowed in this scheme, because they produce higher-derivative couplings in which the non-derivative tachyon appears.
 Moreover, higher-derivative field redefinitions for the metric and dilaton must also be restricted. The dilaton equation of motion is \(\Delta \Phi = 0\), where
\beqa
\Delta \Phi&=&-2R+8\nabla_\mu\Phi\nabla^\mu\Phi-8\nabla_\mu\nabla^\mu\Phi+\frac{1}{6}H_{\mu\nu\alpha}H^{\mu\nu\alpha}+\frac{1}{2}\nabla_\mu T\nabla^\mu T+2V(T)+\cdots\,.
\eeqa
The presence of \(V(T)\) then indicates that the higher-derivative field redefinition of the dilaton produces couplings with the non-derivative tachyon and takes us out of the scheme. Similarly, the equation of motion for the metric is \(\Delta G_{\mu\nu} = 0\), where
\beqa
\Delta G_{\mu\nu}&=&-R_{\mu\nu}-2\nabla_\mu\nabla_\nu\Phi+\frac{1}{4}H_{\mu\alpha\beta}H^{\nu}{}^{\alpha\beta}+\frac{1}{4}\nabla_\mu T\nabla_\nu T-\frac{1}{4}G_{\mu\nu}\Delta\Phi+\cdots\,.
\eeqa
The presence of the tachyon potential in \(\Delta\Phi\) then indicates that the field redefinition of the metric takes us out of the scheme.  It is important to note that if one does not consider the tachyon potential in the above equations of motion, then the corresponding  field redefinitions would treat the tachyon as a massless field and hence would not be consistent with the string theory S-matrix elements in which the mass of the tachyon is taken into account. 

On the other hand, higher-derivative field redefinitions for the Kalb–Ramond field are allowed. The equation of motion for the Kalb–Ramond field is \(\Delta B_{\mu\nu} = 0\), where
\beqa
\Delta B^{\alpha\beta}&=&\frac{1}{2}\nabla_\mu H^{\alpha\beta\mu}-\nabla_\mu\Phi H^{\alpha\beta\mu}+\cdots\,.
\eeqa
Since the tachyon potential does not appear in \(\Delta B_{\mu\nu}\), higher-derivative field redefinitions of the Kalb–Ramond field keep us in the scheme. 
Therefore, to find the basis of couplings at each derivative order in the scheme in which the non-derivative tachyon appears only in the tachyon potential, and in which the basis is consistent with string theory S-matrix elements, one is allowed to use the restricted field redefinition in which one uses total derivatives, Bianchi identities, and only higher-derivative field redefinitions of the Kalb–Ramond field. The resulting basis is then constrained to be invariant under restricted T-duality transformations, which are the Buscher rules supplemented by higher-derivative corrections for all base-space massless fields except the metric and the dilaton. However, since this basis is not the minimal basis, T-duality cannot fix as many coupling constants as in the minimal basis. To use T-duality to fix as many coupling constants as possible in the above non-minimal basis, we employ the following trick: we ignore the tachyon potential in the equations of motion and use unrestricted field redefinitions to find the minimal basis, and then use the unrestricted T-duality constraint on the minimal basis to fix its coupling constants as much as possible. These couplings, however, are not consistent with S-matrix elements. We then use an appropriate field redefinition to change this fixed minimal basis into a new basis that is consistent with the restricted T-duality. This step does not change the number of T-dual multiplets. The resulting couplings must then be consistent with string theory S-matrix elements.

The resulting T-dual couplings contain two parts: one part contains couplings between only massless fields as well as couplings between the tachyon and massless fields, and the other part contains only couplings between the tachyon and massless fields. We call the former couplings the massless sector and the latter couplings the tachyon sector. While the couplings in the tachyon sector cannot be simplified further, the couplings in the massless sector can be simplified greatly. There should exist a  field redefinition that removes all tachyon fields in the massless sector. The reason is that one is allowed to change the couplings in each T-dual set as long as the set remains invariant under the restricted T-duality transformation. Since the tachyon is invariant under T-duality, there are massless couplings that are invariant under the restricted T-duality transformations. These couplings should be physically the same as the couplings in the massless sector. Hence, to find the couplings in the massless sector, we consider the total derivative terms, Bianchi identities, and higher-derivative Kalb–Ramond field redefinitions for the massless fields to find the massless basis. We then use the restricted T-duality transformations to fix their coupling constants. Since the massless basis is not the minimal basis, T-duality cannot fix all coupling constants. However, apart from the coefficients of the Riemann curvature couplings that are invariant under field redefinitions and are the same as the corresponding terms in the minimal T-dual basis, all other parameters are arbitrary and can be set to zero or to any other value that simplifies the final result.

The massless and tachyon sectors at four-derivative order have been found in \cite{Garousi:2026imu}. In this paper, we determine the massless sector and the tachyon sector at six-derivative order. The massless sector, in a particular scheme, contains 83 couplings with fixed coupling constants, while the tachyon sector contains two couplings whose coupling constants must be fixed by the S-matrix method.

To fix the coupling constants in the tachyon sector using the corresponding string-theory S-matrix elements, one requires an expansion of the various physical poles of the S-matrix elements in bosonic string theory. These poles involve tachyon, massless states, and an infinite tower of massive states. It has been speculated that the expansion consistent with a diffeomorphism-invariant spacetime effective action for tachyon and massless fields requires that either the tachyon or the massless pole, along with all massive poles in each channel, should be expanded. This expansion constrains the effective action to contain only an even number of tachyon fields \cite{Garousi:2026nvz}. We will see that the expansion of the sphere-level S-matrix element of two tachyons and two massless vertex operators is fully consistent with the couplings in the massless sector. Moreover, the expansion of the S-matrix element of four tachyon vertex operators fixes one of the couplings in the tachyon sector, while the other coupling involves  six fields and requires analyzing the  six-point function, which we postpone for future work.

The structure of this paper is as follows. In the next section, we show that the existing couplings up to four-derivative order naturally classify into the massless and tachyon sectors. In Section 3, we find the minimal basis at six-derivative order and fix its coupling constants by T-duality. We find that there are three T-dual multiplets: one of them has fixed coupling constants that are invariant under the Buscher rules supplemented by higher-derivative corrections for all base-space massless fields, and the other two have arbitrary coupling constants that are invariant under the Buscher rules without corrections. They form the couplings in the tachyon sector. While the couplings in the tachyon sector are consistent with the S-matrix, the former T-dual multiplet is not. Then, in Section 4, we find the equivalent T-dual multiplet that is consistent with the S-matrix. They form the couplings in the massless sector. We first find the non-minimal basis by using a restricted field redefinition, and then we fix its coupling constants by constraining them to be invariant under the restricted T-duality transformation. There remain some unfixed parameters, which we argue are arbitrary. Using the resulting six-derivative couplings in the massless sector, in Subsection 4.1, we calculate the field theory amplitude of two tachyons and two gravitons or two Kalb–Ramond states. In Subsection 4.2, we show that they are exactly the same as the corresponding terms in the expansion of the sphere-level S-matrix element of two tachyons and two massless NS-NS vertex operators. In Section 5, we show that the coupling in the tachyon sector that has four tachyon fields is consistent with the expansion of the sphere-level S-matrix element of four tachyon vertex operators. These consistency checks confirm both the couplings that we have found in this paper by T-duality and the prescription given in \cite{Garousi:2026nvz} for expanding the S-matrix element in bosonic string theory that involves the tachyon as well as massless fields. In Section 6, we discuss our results.

\section{ Massless and tachyon   sectors}

The effective action of bosonic string theory for tachyon and massless fields, in the most general case, involves all odd and even numbers of tachyons, their partial derivatives, as well as partial derivatives of all massless fields. This is the case if one integrates out the massive states in bosonic string field theory. However, one may use non-covariant field redefinitions to rewrite the effective action in a covariant scheme in which all partial derivatives are replaced by covariant derivatives \cite{Coletti:2003ai,Taylor:2004rh,Michishita:2006dr}. In this case, the analysis of S-matrix elements dictates that the covariant effective action involve only an even number of tachyons \cite{Garousi:2026nvz} . Moreover, using higher-covariant-derivative field redefinitions for the tachyon in the covariant scheme, one can rewrite the covariant couplings in a scheme in which the non-derivative tachyon appears only in the tachyon potential, while all other terms involve covariant derivatives of the tachyon and massless fields \cite{Garousi:2026imu}. Consistency with string theory S-matrix elements  requires that the couplings should not have terms in which \(\nabla_\mu \nabla^\mu\) acts on the tachyon or on derivatives of the tachyon; further higher-derivative field redefinitions or higher-derivative corrections to the T-duality transformation should also be restricted. As we argued in the Introduction, we expect the effective action to be separated into two sectors: the tachyon sector, which involves covariant derivatives of the tachyon and massless fields, and the massless sector, which involves only covariant derivatives of massless fields.
That is,
\beqa
\bS=\sum_{n=-1}^{\infty}\alpha'^n\bS^{(n)}&;& \bS^{(n)}=\frac{2}{\kappa^2}\int d^{26}x\sqrt{-G}\,e^{-2\Phi}\big[\cL_T^{(n)}+\cL_M^{(n)}\Big]\,,\labell{Sn1}
\eeqa 
where \(\mathcal{L}_T^{(n)}\) contains couplings in the tachyon sector at order \(\alpha'^n\), and \(\mathcal{L}_M^{(n)}\) contains couplings in the massless sector at order \(\alpha'^n\).

At order \(\alpha'^{-1}\), or in the non-derivative part, the massless sector is zero and the tachyon sector contains the tachyon potential, i.e., \(\mathcal{L}_T^{(-1)} = -V(T)\). The tachyon potential is invariant under T-duality; hence, this term can only be determined by the S-matrix method. The tachyon mass gives the quadratic part of this potential, which indicates that the tachyon potential at \(T=0\) is unstable. The expansion of the sphere-level S-matrix element of four tachyon vertex operators fixes the quartic part of this potential to be \cite{Garousi:2026imu}
\beqa
V(T)&=&-\frac{1}{2!}T^2+\frac{1}{4!}T^4+\cdots\,,\labell{VT1}
\eeqa
where the dots represent terms at order \(T^6\) and higher, which require analyzing the S-matrix element of six tachyon vertex operators and higher. The sign of the quartic term indicates that the potential in fact has a minimum at \(T \neq 0\), as well as a maximum at \(T = 0\).

At two-derivative order, the tachyon and massless sectors are
\beqa
 \cL_T^{(0)}&=&- \frac{1}{4}\partial_\mu T \partial^\mu T\,,\nn\\
\cL_M^{(0)}&=&R + 4\partial_\mu \Phi \partial^\mu \Phi - \frac{1}{12}H^2\,.\labell{S0}
\eeqa
The above couplings are  invariant under the Buscher rules. The simplest form of these rules appears if one considers the following reduction for the tachyon and massless fields \cite{Maharana:1992my}:
\beqa  
G_{\mu\nu} = \left(\matrix{\bg_{ab} + e^{\varphi} g_{a} g_{b} &  e^{\varphi} g_{a} \cr e^{\varphi} g_{b} & e^{\varphi} &}\!\!\!\!\!\right), \quad  
B_{\mu\nu} = \left(\matrix{\bb_{ab} + b_{[a} g_{b]} & b_{a} \cr -b_{b} & 0 &}\!\!\!\!\!\right), \quad  
\Phi = \bar{\phi} + \varphi/4,  \quad T=\bar{T}\,,
\labell{reduction1}  
\eeqa  
where indices \(a,b\) denote directions orthogonal to the Killing coordinate \(y\). In the above reduction, \(\bar{g}_{ab}\) is the \((D-1)\)-dimensional base-space metric, \(\bar{b}_{ab}\) is an antisymmetric tensor, \(\bar{\phi}\) is the dilaton, \(\bar{T}\) is the tachyon, \(\varphi\) is a scalar, and \(g_a, b_a\) are two vectors. The Buscher rules \cite{Buscher:1987sk,Rocek:1991ps} then take the following form:
\beqa  
&&\bg_{ab}' = \bg_{ab}\,, \quad 
\bar{T}'=\bar{T}\,, \quad 
\bar{\phi}' = \bar{\phi}\,, \nn\\
&&\varphi' = -\varphi\,, \quad 
g_a' = b_a\,, \quad 
b_a' = g_a\,, \quad \bar{H}_{abc}' = \bar{H}_{abc}\,,
\labell{flac1}  
\eeqa  
where $\bar{H}_{abc}$ is torsion in the base space \cite{Kaloper:1997ux}. Under the reduction \reef{reduction1}, one finds \(\sqrt{-G}\,e^{-2\Phi} = \sqrt{-\bar{g}}\,e^{-2\bar{\phi}}\). Hence, the circular reduction of the tachyon sector is invariant under T-duality, and the reduction of the massless sector is separately invariant, up to a total derivative term in the base space, under the above Buscher rules \cite{Garousi:2019wgz}. For consistency of the higher-derivative couplings with T-duality and with S-matrix, the transformations in the second line above must receive higher-derivative corrections, while the transformations in the first line above should not receive corrections.

The tachyon and massless sectors at order \(\alpha'\) have been found in \cite{Garousi:2026imu} by the T-duality and S-matrix methods to be
 \beqa
  \cL_T^{(1)}&=&\frac{1}{32}(\nabla_\alpha T\nabla^\alpha T)^2\,,\nn\\
\cL_M^{(1)}&=&\frac{1}{4}\Big[ \frac{1}{24} H_{\alpha 
}{}^{\delta \epsilon } H^{\alpha \beta \gamma } H_{\beta 
\delta }{}^{\varepsilon } H_{\gamma \epsilon \varepsilon } -  
\frac{1}{8} H_{\alpha \beta }{}^{\delta } H^{\alpha \beta 
\gamma } H_{\gamma }{}^{\epsilon \varepsilon } H_{\delta 
\epsilon \varepsilon } \nn\\&&+ \frac{1}{144} H_{\alpha \beta \gamma 
} H^{\alpha \beta \gamma } H_{\delta \epsilon \varepsilon } H^{
\delta \epsilon \varepsilon } + H_{\alpha }{}^{\gamma \delta } 
H_{\beta \gamma \delta } R^{\alpha \beta } - 4 
R_{\alpha \beta } R^{\alpha \beta } -  
\frac{1}{6} H_{\alpha \beta \gamma } H^{\alpha \beta \gamma } 
R + R^2 \nn\\&&+ R_{\alpha \beta \gamma 
\delta } R^{\alpha \beta \gamma \delta } -  
\frac{1}{2} H_{\alpha }{}^{\delta \epsilon } H^{\alpha \beta 
\gamma } R_{\beta \gamma \delta \epsilon } -  
\frac{2}{3} H_{\beta \gamma \delta } H^{\beta \gamma \delta } 
\nabla_{\alpha }\nabla^{\alpha }\Phi + \frac{2}{3} H_{\beta 
\gamma \delta } H^{\beta \gamma \delta } \nabla_{\alpha }\Phi 
\nabla^{\alpha }\Phi \nn\\&&+ 8 R \nabla_{\alpha }\Phi 
\nabla^{\alpha }\Phi - 16 R_{\alpha \beta } 
\nabla^{\alpha }\Phi \nabla^{\beta }\Phi + 16 \nabla_{\alpha 
}\Phi \nabla^{\alpha }\Phi \nabla_{\beta }\Phi \nabla^{\beta 
}\Phi - 32 \nabla^{\alpha }\Phi \nabla_{\beta }\nabla_{\alpha 
}\Phi \nabla^{\beta }\Phi \nn\\&&+ 2 H_{\alpha }{}^{\gamma \delta } 
H_{\beta \gamma \delta } \nabla^{\beta }\nabla^{\alpha }\Phi\Big]\,.\labell{fourmax3}
\eeqa
The massless sector is the effective action in the Meissner scheme \cite{Meissner:1996sa}. The tachyon sector is obviously invariant under T-duality, and the massless sector is invariant under the Buscher rules, provided the corrections to the transformations in the second line of \reef{flac1} are

\beqa
 \Delta\bg_{a}^{(1)}&=&\frac{1}{4}\Big( \bH_{abc} V^{bc}+2{e^{-\vp}} W_{ab}\prt^b\vp\Big)\,,\nn\\
  \Delta\bb_{a}^{(1)}&=&\frac{1}{4}\Big( -\bH_{abc} W^{bc}+2{e^{\vp}} V_{ab}\prt^b\vp\Big)\,,\nn\\
  \Delta\vp^{(1)}&=&\frac{1}{4}\Big( e^{\vp} V_{ab} V^{ab}  + {e^{-\vp}} W_{ab} W^{ab} + 2 \prt_{a}\vp \prt^{a}\vp\Big)\,,\nn\\
\Delta\bar{B}_{ab}^{(1)}&=& - V_{[b}{}^{c} W_{a]c}\,.\labell{d1}
\eeqa
The field strength of the last term appears in the correction to the three-form $\bH$. These transformations were first found in \cite{Kaloper:1997ux}. It has been shown in \cite{Garousi:2026imu} that the S-matrix element of two tachyons and two gravitons, calculated from the tachyon potential \reef{VT1}, the tachyon sector in \reef{S0} and the massless sector in \reef{fourmax3}, is produced exactly by the second term in the expansion of the string-theory S-matrix element of two tachyon and two graviton vertex operators. Moreover, the second term in the expansion of the S-matrix element of four tachyon vertex operators produces the quartic term in the tachyon potential \reef{VT1}, as well as the derivative coupling in the tachyon sector in \reef{fourmax3}.

\section{Minimal basis at order $\alpha'^2$}

To use T-duality to find the effective action at six-derivative order for massless and tachyon fields, one needs to use field redefinitions to find the independent couplings. As we argued in the Introduction, in the scheme in which the non-derivative tachyon appears only in the tachyon potential, consistency with string theory S-matrix elements requires the field redefinitions to be restricted so that they include no corrections to the metric, dilaton, and tachyon, and the corresponding corrections to the Buscher rules should not include corrections to the base-space metric, dilaton, and tachyon. However, the basis found using this restricted field redefinition cannot be fully fixed by the corresponding restricted T-duality. There remain many coupling constants that require S-matrix calculations to fix them. On the other hand, the minimal basis, in which the tachyon potential in the equations of motion is ignored and field redefinitions of all massless fields are employed, can be maximally fixed by Buscher rules in which higher-derivative corrections for all base-space massless fields are used. There remain a few coupling constants in the T-dual couplings, each corresponding to one T-dual multiplet. However, such T-dual couplings are not consistent with S-matrix elements. We therefore use the trick of first finding the T-dual couplings in the minimal basis, and then using an appropriate field redefinition to rewrite them such that they become invariant under the restricted T-duality transformation. As a result, the number of T-dual multiplets is the same as in the minimal basis, and the new couplings should be consistent with the S-matrix.

To find the minimal couplings involving the massless and tachyon fields, we first use the xAct package \cite{Nutma:2013zea} to find all couplings at order \(\alpha'^2\) that involve covariant derivatives of the tachyon and massless fields, as well as Riemann, Ricci, and scalar curvatures. There are 969 such couplings. Following \cite{Garousi:2019cdn}, and using total derivatives, Bianchi identities, and field redefinitions of all massless fields, we find that they produce 110 relations among the 969 coupling constants. We should set some of these parameters to zero such that there remain 110 relations for the parameters. Among the parameters that we set to zero, we choose all coefficients of couplings that include the operator \(\nabla_\mu \nabla^\mu\) acting on the tachyon or on derivatives of the tachyon. In S-matrix elements, whenever there is a factor of \(p_i^2\), where \(p_i^\mu\) is the momentum of the \(i^{\rm th}\)-tachyon vertex operator, one must replace it with minus the square of the tachyon mass. In spacetime, this means that \(\nabla_\mu \nabla^\mu T(\cdots)\) should be equivalent to \(-2T(\cdots)\). Since the latter coupling does not exist in our scheme, consistency with the S-matrix requires that the former coupling should not exist in the scheme either. It turns out that when we set the coefficients of such terms to zero, there still remain 110 relations among the remaining parameters. Hence, removing such couplings is merely a choice of scheme; it does not restrict the independent couplings in the minimal basis. We write the minimal basis in the following scheme:
\beqa
 &\!\!\!\!\!\!\!&c_{1} 
   H_{\alpha  }{}^{\delta  \epsilon  } H^{\alpha  \beta  
\gamma  } H_{\beta  \delta  }{}^{\varepsilon  } H_{\gamma  }{}^{
\mu \zeta  } H_{\epsilon\mu}{}^{\eta  } 
H_{\varepsilon  \zeta  \eta  } +c_{2} 
   H_{\alpha  \beta  }{}^{\delta  } H^{\alpha  \beta  \gamma  
} H_{\gamma  }{}^{\epsilon  \varepsilon  } H_{\delta  
}{}^{\mu \zeta  } H_{\epsilon  \varepsilon  }{}^{\eta  
} H_{\mu \zeta  \eta  }  \nn\\&\!\!\!\!\!\!\!&+c_{3} 
   H_{\alpha  \beta  }{}^{\delta  } H^{\alpha  \beta  \gamma  
} H_{\gamma  }{}^{\epsilon  \varepsilon  } H_{\delta  \epsilon  
}{}^{\mu } H_{\varepsilon  }{}^{\zeta  \eta  } 
H_{\mu \zeta  \eta  } +c_{4} 
  R_{\alpha  }{}^{\epsilon  }{}_{\gamma  
}{}^{\varepsilon  }R^{\alpha  \beta  \gamma  \delta  } 
R_{\beta  \varepsilon  \delta  \epsilon  } + 
 c_{5} 
  R_{\alpha  \beta  }{}^{\epsilon  \varepsilon  } 
R^{\alpha  \beta  \gamma  \delta  }R_{\gamma  
\epsilon  \delta  \varepsilon  }  \nn\\&\!\!\!\!\!\!\!&+c_{6} 
   H_{\alpha  }{}^{\delta  \epsilon  } H^{\alpha  \beta  
\gamma  } H_{\beta  }{}^{\varepsilon\mu} H_{\delta  
\varepsilon  }{}^{\zeta  }R_{\gamma  \epsilon  
\mu \zeta  } +c_{7} 
   H_{\alpha  }{}^{\delta  \epsilon  } H^{\alpha  \beta  
\gamma  }R_{\beta  }{}^{\varepsilon  }{}_{\delta  }{}^{
\mu }R_{\gamma  \varepsilon  \epsilon  
\mu } +c_{8} 
   H_{\alpha  \beta  }{}^{\delta  } H^{\alpha  \beta  \gamma  
} H_{\epsilon  \varepsilon  }{}^{\zeta  } H^{\epsilon  
\varepsilon\mu}R_{\gamma\mu
\delta  \zeta  }  \nn\\&\!\!\!\!\!\!\!&+c_{9} 
   H^{\alpha  \beta  \gamma  } H^{\delta  \epsilon  
\varepsilon  }R_{\alpha  \beta  \delta  
}{}^{\mu }R_{\gamma\mu\epsilon  
\varepsilon  } +c_{10} 
   H_{\alpha  }{}^{\delta  \epsilon  } H^{\alpha  \beta  
\gamma  }R_{\beta  }{}^{\varepsilon  }{}_{\delta  }{}^{
\mu }R_{\gamma\mu\epsilon  
\varepsilon  } +c_{11} 
   H_{\alpha  }{}^{\delta  \epsilon  } H^{\alpha  \beta  
\gamma  }R_{\beta  }{}^{\varepsilon  }{}_{\gamma  }{}^{
\mu }R_{\delta  \varepsilon  \epsilon  
\mu }  \nn\\&\!\!\!\!\!\!\!&+c_{12} 
   H_{\alpha  \beta  }{}^{\delta  } H^{\alpha  \beta  \gamma  
}R_{\gamma  }{}^{\epsilon  \varepsilon\mu} 
R_{\delta  \varepsilon  \epsilon\mu} + 
 c_{13} 
   H_{\alpha  \beta  }{}^{\delta  } H^{\alpha  \beta  \gamma  
} H_{\gamma  }{}^{\epsilon  \varepsilon  } H_{\epsilon  
}{}^{\mu \zeta  }R_{\delta\mu
\varepsilon  \zeta  } +c_{14} 
   H_{\alpha  }{}^{\delta  \epsilon  } H^{\alpha  \beta  
\gamma  } H_{\beta  \delta  }{}^{\varepsilon  } H_{\gamma  }{}^{
\mu \zeta  }R_{\epsilon\mu
\varepsilon  \zeta  }  \nn\\&\!\!\!\!\!\!\!&+c_{15} 
   H_{\alpha  \beta  }{}^{\delta  } H^{\alpha  \beta  \gamma  
} H_{\gamma  }{}^{\epsilon  \varepsilon  } H_{\delta  
}{}^{\mu \zeta  }R_{\epsilon\mu
\varepsilon  \zeta  } +c_{16} 
   H_{\beta  }{}^{\epsilon  \varepsilon  } H^{\beta  \gamma  
\delta  } H_{\gamma  \epsilon  }{}^{\mu } H_{\delta  
\varepsilon\mu} \nabla_{\alpha  }T \nabla^{\alpha  
}T +c_{17} 
  R_{\beta  \delta  \gamma  \epsilon  } 
R^{\beta  \gamma  \delta  \epsilon  } \nabla_{\alpha  
}T \nabla^{\alpha  }T  \nn\\&\!\!\!\!\!\!\!&+c_{18} 
   H_{\beta  }{}^{\epsilon  \varepsilon  } H^{\beta  \gamma  
\delta  }R_{\gamma  \epsilon  \delta  \varepsilon  } 
\nabla_{\alpha  }T \nabla^{\alpha  }T +c_{19} 
   H_{\beta  }{}^{\epsilon  \varepsilon  } H^{\beta  \gamma  
\delta  } H_{\gamma  \epsilon  }{}^{\mu } H_{\delta  
\varepsilon\mu} \nabla_{\alpha  }\Phi  
\nabla^{\alpha  }\Phi  +c_{20} 
  R_{\beta  \gamma  \delta  \epsilon  } 
R^{\beta  \gamma  \delta  \epsilon  } \nabla_{\alpha  
}\Phi  \nabla^{\alpha  }\Phi   \nn\\&\!\!\!\!\!\!\!&+c_{21} 
   H_{\beta  }{}^{\epsilon  \varepsilon  } H^{\beta  \gamma  
\delta  }R_{\gamma  \epsilon  \delta  \varepsilon  } 
\nabla_{\alpha  }\Phi  \nabla^{\alpha  }\Phi  + 
 c_{22} 
   H_{\alpha  }{}^{\gamma  \delta  } H_{\beta  }{}^{\epsilon  
\varepsilon  } H_{\gamma  \epsilon  }{}^{\mu } 
H_{\delta  \varepsilon\mu} \nabla^{\alpha  }T 
\nabla^{\beta  }T +c_{23} 
   H_{\alpha  }{}^{\gamma  \delta  } H_{\beta  }{}^{\epsilon  
\varepsilon  }R_{\gamma  \epsilon  \delta  \varepsilon 
 } \nabla^{\alpha  }T \nabla^{\beta  }T  \nn\\&\!\!\!\!\!\!\!&+c_{24} 
   H_{\alpha  }{}^{\gamma  \delta  } H_{\beta  }{}^{\epsilon  
\varepsilon  } H_{\gamma  \epsilon  }{}^{\mu } 
H_{\delta  \varepsilon\mu} \nabla^{\alpha  }\Phi  
\nabla^{\beta  }\Phi  +c_{25} 
  R_{\alpha  }{}^{\gamma  \delta  \epsilon  } 
R_{\beta  \delta  \gamma  \epsilon  } \nabla^{\alpha  
}\Phi  \nabla^{\beta  }\Phi  +c_{26} 
   H_{\alpha  }{}^{\gamma  \delta  } H_{\beta  }{}^{\epsilon  
\varepsilon  }R_{\gamma  \epsilon  \delta  \varepsilon 
 } \nabla^{\alpha  }\Phi  \nabla^{\beta  }\Phi  \nn\\&\!\!\!\!\!\!\!& + 
 c_{27} 
   H_{\gamma  \delta  \epsilon  } H^{\gamma  \delta  \epsilon 
 } \nabla^{\alpha  }\Phi  \nabla_{\beta  }\nabla_{\alpha  
}\Phi  \nabla^{\beta  }\Phi  +c_{28} 
   H_{\alpha  }{}^{\gamma  \delta  } H_{\beta  }{}^{\epsilon  
\varepsilon  } H_{\gamma  \epsilon  }{}^{\mu } 
H_{\delta  \varepsilon\mu} \nabla^{\beta  
}\nabla^{\alpha  }\Phi  +c_{29} 
   H_{\alpha  }{}^{\gamma  \delta  } H_{\beta  }{}^{\epsilon  
\varepsilon  } H_{\gamma  \delta  }{}^{\mu } 
H_{\epsilon  \varepsilon\mu} \nabla^{\beta  
}\nabla^{\alpha  }\Phi   \nn\\&\!\!\!\!\!\!\!&+c_{30} 
   H_{\alpha  }{}^{\gamma  \delta  } H_{\beta  \gamma  
}{}^{\epsilon  } H_{\delta  }{}^{\varepsilon\mu} H_{
\epsilon  \varepsilon\mu} \nabla^{\beta  }\nabla^{
\alpha  }\Phi  +c_{31} 
   H_{\gamma  \delta  }{}^{\varepsilon  } H^{\gamma  \delta  
\epsilon  }R_{\alpha  \epsilon  \beta  \varepsilon  } 
\nabla^{\beta  }\nabla^{\alpha  }\Phi  +c_{32} 
  R_{\alpha  }{}^{\gamma  \delta  \epsilon  } 
R_{\beta  \delta  \gamma  \epsilon  } \nabla^{\beta  }
\nabla^{\alpha  }\Phi  \nn\\&\!\!\!\!\!\!\!& +c_{33} 
   H_{\alpha  }{}^{\gamma  \delta  } H_{\gamma  }{}^{\epsilon  
\varepsilon  }R_{\beta  \epsilon  \delta  \varepsilon  
} \nabla^{\beta  }\nabla^{\alpha  }\Phi  +c_{34} 
   H_{\alpha  }{}^{\gamma  \delta  } H_{\beta  }{}^{\epsilon  
\varepsilon  }R_{\gamma  \epsilon  \delta  \varepsilon 
 } \nabla^{\beta  }\nabla^{\alpha  }\Phi  +c_{35} 
   \nabla_{\alpha  }T \nabla^{\alpha  }T \nabla_{\beta  }T 
\nabla^{\beta  }T \nabla_{\gamma  }T \nabla^{\gamma  }T \nn\\&\!\!\!\!\!\!\!& + 
 c_{36} 
   \nabla_{\alpha  }T \nabla^{\alpha  }T \nabla_{\beta  
}\Phi  \nabla^{\beta  }T \nabla_{\gamma  }\Phi  
\nabla^{\gamma  }T +c_{37} 
   \nabla_{\alpha  }\Phi  \nabla^{\alpha  }T \nabla^{\beta  
}T \nabla_{\gamma  }\nabla_{\beta  }T \nabla^{\gamma  }T \nn\\&\!\!\!\!\!\!\!& + 
 c_{38} 
   \nabla_{\alpha  }T \nabla^{\alpha  }T \nabla^{\beta  }T 
\nabla_{\gamma  }\nabla_{\beta  }\Phi  \nabla^{\gamma  }T +
  c_{39} 
   \nabla_{\alpha  }\nabla_{\gamma  }\nabla_{\beta  }T 
\nabla^{\alpha  }T \nabla^{\beta  }\Phi  \nabla^{\gamma  
}\Phi   \nn\\&\!\!\!\!\!\!\!&+c_{40} 
   H_{\alpha  }{}^{\delta  \epsilon  } \nabla^{\alpha  }\Phi  
\nabla^{\beta  }\Phi  \nabla_{\gamma  }H_{\beta  \delta  
\epsilon  } \nabla^{\gamma  }\Phi  +c_{41} 
   H_{\alpha  }{}^{\delta  \epsilon  } H_{\beta  \delta  
\epsilon  } \nabla^{\alpha  }T \nabla^{\beta  }T 
\nabla_{\gamma  }\Phi  \nabla^{\gamma  }\Phi  \nn\\&\!\!\!\!\!\!\!& + 
 c_{42} 
   \nabla_{\alpha  }T \nabla^{\alpha  }T \nabla_{\beta  }T 
\nabla^{\beta  }T \nabla_{\gamma  }\Phi  \nabla^{\gamma  
}\Phi  +c_{43} 
   \nabla_{\alpha  }\Phi  \nabla^{\alpha  }T \nabla_{\beta  
}\Phi  \nabla^{\beta  }T \nabla_{\gamma  }\Phi  
\nabla^{\gamma  }\Phi  \labell{Lc}\\&\!\!\!\!\!\!\!& +c_{44} 
   \nabla^{\alpha  }T \nabla_{\beta  }\nabla_{\alpha  }\Phi  
\nabla^{\beta  }T \nabla_{\gamma  }\Phi  \nabla^{\gamma  
}\Phi  +c_{45} 
   \nabla_{\alpha  }T \nabla^{\alpha  }T \nabla_{\beta  
}\Phi  \nabla^{\beta  }\Phi  \nabla_{\gamma  }\Phi  
\nabla^{\gamma  }\Phi  \nn\\&\!\!\!\!\!\!\!& +c_{46} 
   \nabla_{\alpha  }\Phi  \nabla^{\alpha  }\Phi  
\nabla_{\beta  }\Phi  \nabla^{\beta  }\Phi  \nabla_{\gamma  }
\Phi  \nabla^{\gamma  }\Phi  +c_{47} 
   \nabla_{\alpha  }\Phi  \nabla^{\alpha  }T \nabla^{\beta  
}\Phi  \nabla_{\gamma  }\nabla_{\beta  }T \nabla^{\gamma  
}\Phi  \nn\\&\!\!\!\!\!\!\!& +c_{48} 
   \nabla_{\alpha  }\Phi  \nabla^{\alpha  }T \nabla^{\beta  
}T \nabla_{\gamma  }\nabla_{\beta  }\Phi  \nabla^{\gamma  
}\Phi  +c_{49} 
   \nabla_{\alpha  }T \nabla^{\alpha  }T \nabla^{\beta  
}\Phi  \nabla_{\gamma  }\nabla_{\beta  }\Phi  \nabla^{\gamma  
}\Phi   \nn\\&\!\!\!\!\!\!\!&+c_{50} 
   \nabla_{\alpha  }\Phi  \nabla^{\alpha  }\Phi  
\nabla^{\beta  }\Phi  \nabla_{\gamma  }\nabla_{\beta  }\Phi  
\nabla^{\gamma  }\Phi  +c_{51} 
   \nabla^{\alpha  }T \nabla^{\beta  }\Phi  \nabla_{\gamma  
}\nabla_{\beta  }\nabla_{\alpha  }T \nabla^{\gamma  }\Phi  \nn\\&\!\!\!\!\!\!\!& + 
c_{52} 
   H_{\beta  }{}^{\delta  \epsilon  } H_{\gamma  \delta  
\epsilon  } \nabla^{\alpha  }T \nabla^{\beta  }\Phi  
\nabla^{\gamma  }\nabla_{\alpha  }T +c_{53} 
   H_{\beta  }{}^{\delta  \epsilon  } H_{\gamma  \delta  
\epsilon  } \nabla^{\beta  }\nabla^{\alpha  }T \nabla^{\gamma  
}\nabla_{\alpha  }T +c_{54} 
   \nabla^{\alpha  }T \nabla^{\beta  }T \nabla_{\gamma  
}\nabla_{\beta  }T \nabla^{\gamma  }\nabla_{\alpha  }T  \nn\\&\!\!\!\!\!\!\!&+ 
 c_{55} 
   \nabla^{\alpha  }\Phi  \nabla^{\beta  }\Phi  
\nabla_{\gamma  }\nabla_{\beta  }T \nabla^{\gamma  
}\nabla_{\alpha  }T +c_{56} 
   \nabla^{\alpha  }T \nabla^{\beta  }\Phi  \nabla_{\gamma  
}\nabla_{\beta  }\Phi  \nabla^{\gamma  }\nabla_{\alpha  }T + 
c_{57} 
   \nabla^{\beta  }\nabla^{\alpha  }T \nabla_{\gamma  
}\nabla_{\beta  }\Phi  \nabla^{\gamma  }\nabla_{\alpha  }T \nn\\&\!\!\!\!\!\!\!& + 
c_{58} 
   H_{\beta  }{}^{\delta  \epsilon  } H_{\gamma  \delta  
\epsilon  } \nabla^{\alpha  }T \nabla^{\beta  }T 
\nabla^{\gamma  }\nabla_{\alpha  }\Phi  +c_{59} 
   H_{\beta  }{}^{\delta  \epsilon  } H_{\gamma  \delta  
\epsilon  } \nabla^{\alpha  }\Phi  \nabla^{\beta  }\Phi  
\nabla^{\gamma  }\nabla_{\alpha  }\Phi  +c_{60} 
   H_{\beta  }{}^{\delta  \epsilon  } H_{\gamma  \delta  
\epsilon  } \nabla^{\beta  }\nabla^{\alpha  }\Phi  
\nabla^{\gamma  }\nabla_{\alpha  }\Phi  \nn\\&\!\!\!\!\!\!\!& +c_{61} 
   \nabla^{\alpha  }T \nabla^{\beta  }T \nabla_{\gamma  
}\nabla_{\beta  }\Phi  \nabla^{\gamma  }\nabla_{\alpha  }\Phi 
 +c_{62} 
   \nabla^{\alpha  }\Phi  \nabla^{\beta  }\Phi  
\nabla_{\gamma  }\nabla_{\beta  }\Phi  \nabla^{\gamma  
}\nabla_{\alpha  }\Phi  +c_{63} 
   \nabla^{\beta  }\nabla^{\alpha  }\Phi  \nabla_{\gamma  
}\nabla_{\beta  }\Phi  \nabla^{\gamma  }\nabla_{\alpha  }\Phi 
  \nn\\&\!\!\!\!\!\!\!&+c_{64} 
   \nabla^{\alpha  }T \nabla^{\beta  }\Phi  \nabla_{\gamma  
}\nabla_{\alpha  }\Phi  \nabla^{\gamma  }\nabla_{\beta  }T + 
c_{65} 
   H_{\beta  }{}^{\delta  \epsilon  } H_{\gamma  \delta  
\epsilon  } \nabla_{\alpha  }\Phi  \nabla^{\alpha  }T \nabla^{
\gamma  }\nabla^{\beta  }T +c_{66} 
   \nabla_{\alpha  }\nabla_{\gamma  }\nabla_{\beta  }T 
\nabla^{\alpha  }\Phi  \nabla^{\gamma  }\nabla^{\beta  }T \nn\\&\!\!\!\!\!\!\!& +
  c_{67} 
   H_{\beta  }{}^{\delta  \epsilon  } \nabla^{\alpha  }T 
\nabla_{\gamma  }H_{\alpha  \delta  \epsilon  } \nabla^{\gamma 
 }\nabla^{\beta  }T +c_{68} 
   H_{\alpha  }{}^{\delta  \epsilon  } \nabla^{\alpha  }T 
\nabla_{\gamma  }H_{\beta  \delta  \epsilon  } \nabla^{\gamma  
}\nabla^{\beta  }T +c_{69} 
   \nabla_{\alpha  }\Phi  \nabla^{\alpha  }\Phi  
\nabla_{\gamma  }\nabla_{\beta  }T \nabla^{\gamma  
}\nabla^{\beta  }T  \nn\\&\!\!\!\!\!\!\!&+c_{70} 
   \nabla_{\alpha  }\Phi  \nabla^{\alpha  }T \nabla_{\gamma  
}\nabla_{\beta  }\Phi  \nabla^{\gamma  }\nabla^{\beta  }T +
  c_{71} 
   \nabla^{\alpha  }\Phi  \nabla_{\gamma  }\nabla_{\beta  
}\nabla_{\alpha  }T \nabla^{\gamma  }\nabla^{\beta  }T + 
 c_{72} 
   H_{\beta  }{}^{\delta  \epsilon  } H_{\gamma  \delta  
\epsilon  } \nabla_{\alpha  }T \nabla^{\alpha  }T 
\nabla^{\gamma  }\nabla^{\beta  }\Phi  \nn\\&\!\!\!\!\!\!\!& +c_{73} 
   \nabla_{\alpha  }\nabla_{\gamma  }\nabla_{\beta  }T 
\nabla^{\alpha  }T \nabla^{\gamma  }\nabla^{\beta  }\Phi  +
  c_{74} 
   H_{\beta  }{}^{\delta  \epsilon  } H_{\gamma  \delta  
\epsilon  } \nabla_{\alpha  }\Phi  \nabla^{\alpha  }\Phi  
\nabla^{\gamma  }\nabla^{\beta  }\Phi  +c_{75} 
   H_{\alpha  }{}^{\delta  \epsilon  } \nabla^{\alpha  }\Phi  
\nabla_{\gamma  }H_{\beta  \delta  \epsilon  } \nabla^{\gamma  
}\nabla^{\beta  }\Phi   \nn\\&\!\!\!\!\!\!\!&+c_{76} 
   \nabla_{\alpha  }T \nabla^{\alpha  }T \nabla_{\gamma  
}\nabla_{\beta  }\Phi  \nabla^{\gamma  }\nabla^{\beta  }\Phi  
+c_{77} 
   \nabla_{\alpha  }\Phi  \nabla^{\alpha  }\Phi  
\nabla_{\gamma  }\nabla_{\beta  }\Phi  \nabla^{\gamma  
}\nabla^{\beta  }\Phi  +c_{78} 
   \nabla^{\alpha  }T \nabla_{\gamma  }\nabla_{\beta  
}\nabla_{\alpha  }T \nabla^{\gamma  }\nabla^{\beta  }\Phi \nn\\&\!\!\!\!\!\!\!&  + 
c_{79} 
   \nabla_{\beta  }\nabla_{\gamma  }\nabla_{\alpha  }T 
\nabla^{\gamma  }\nabla^{\beta  }\nabla^{\alpha  }T + 
 c_{80} 
   \nabla_{\gamma  }\nabla_{\beta  }\nabla_{\alpha  }T 
\nabla^{\gamma  }\nabla^{\beta  }\nabla^{\alpha  }T + 
 c_{81} 
   H_{\alpha  \gamma  }{}^{\epsilon  } H_{\beta  \delta  
\epsilon  } \nabla^{\alpha  }T \nabla^{\beta  }T 
\nabla^{\gamma  }\Phi  \nabla^{\delta  }\Phi  \nn\\&\!\!\!\!\!\!\!& + 
 c_{82} 
   H_{\alpha  \gamma  }{}^{\epsilon  } H_{\beta  \delta  
\epsilon  } \nabla^{\alpha  }T \nabla^{\beta  }\Phi  
\nabla^{\delta  }\nabla^{\gamma  }T +c_{83} 
   H_{\alpha  \gamma  }{}^{\epsilon  } H_{\beta  \delta  
\epsilon  } \nabla^{\beta  }\nabla^{\alpha  }T \nabla^{\delta  
}\nabla^{\gamma  }T +c_{84} 
  R_{\alpha  \gamma  \beta  \delta  } \nabla^{\beta 
 }\nabla^{\alpha  }T \nabla^{\delta  }\nabla^{\gamma  }T  \nn\\&\!\!\!\!\!\!\!&+ 
 c_{85} 
   H_{\alpha  \gamma  }{}^{\epsilon  } H_{\beta  \delta  
\epsilon  } \nabla^{\alpha  }T \nabla^{\beta  }T 
\nabla^{\delta  }\nabla^{\gamma  }\Phi  +c_{86} 
   H_{\alpha  \gamma  }{}^{\epsilon  } H_{\beta  \delta  
\epsilon  } \nabla^{\alpha  }\Phi  \nabla^{\beta  }\Phi  
\nabla^{\delta  }\nabla^{\gamma  }\Phi  +c_{87} 
  R_{\alpha  \gamma  \beta  \delta  } 
\nabla^{\alpha  }\Phi  \nabla^{\beta  }\Phi  \nabla^{\delta  
}\nabla^{\gamma  }\Phi  \nn\\&\!\!\!\!\!\!\!& +c_{88} 
   H_{\alpha  \gamma  }{}^{\epsilon  } H_{\beta  \delta  
\epsilon  } \nabla^{\beta  }\nabla^{\alpha  }\Phi  
\nabla^{\delta  }\nabla^{\gamma  }\Phi  +c_{89} 
  R_{\alpha  \gamma  \beta  \delta  } \nabla^{\beta 
 }\nabla^{\alpha  }\Phi  \nabla^{\delta  }\nabla^{\gamma  
}\Phi  +c_{90} 
   H_{\beta  }{}^{\delta  \epsilon  } \nabla^{\alpha  }T 
\nabla^{\gamma  }\nabla^{\beta  }T \nabla_{\epsilon  
}H_{\alpha  \gamma  \delta  }  \nn\\&\!\!\!\!\!\!\!&+c_{91} 
   H_{\beta  }{}^{\delta  \epsilon  } \nabla^{\alpha  }\Phi  
\nabla^{\gamma  }\nabla^{\beta  }\Phi  \nabla_{\epsilon  
}H_{\alpha  \gamma  \delta  } +c_{92} 
   H^{\gamma  \delta  \epsilon  } \nabla^{\alpha  }\Phi  
\nabla^{\beta  }\nabla_{\alpha  }\Phi  \nabla_{\epsilon  
}H_{\beta  \gamma  \delta  } +c_{93} 
   \nabla^{\alpha  }T \nabla^{\beta  }T \nabla_{\delta  
}H_{\beta  \gamma  \epsilon  } \nabla^{\epsilon  }H_{\alpha  
}{}^{\gamma  \delta  } \nn\\&\!\!\!\!\!\!\!& +c_{94} 
   \nabla^{\alpha  }\Phi  \nabla^{\beta  }\Phi  
\nabla_{\delta  }H_{\beta  \gamma  \epsilon  } 
\nabla^{\epsilon  }H_{\alpha  }{}^{\gamma  \delta  } + 
 c_{95} 
   \nabla^{\beta  }\nabla^{\alpha  }\Phi  \nabla_{\delta  
}H_{\beta  \gamma  \epsilon  } \nabla^{\epsilon  }H_{\alpha  
}{}^{\gamma  \delta  } +c_{96} 
   \nabla^{\alpha  }T \nabla^{\beta  }T \nabla_{\epsilon  
}H_{\beta  \gamma  \delta  } \nabla^{\epsilon  }H_{\alpha  
}{}^{\gamma  \delta  }  \nn\\&\!\!\!\!\!\!\!&+c_{97} 
   \nabla^{\alpha  }\Phi  \nabla^{\beta  }\Phi  
\nabla_{\epsilon  }H_{\beta  \gamma  \delta  } 
\nabla^{\epsilon  }H_{\alpha  }{}^{\gamma  \delta  } + 
 c_{98} 
   \nabla^{\beta  }\nabla^{\alpha  }\Phi  \nabla_{\epsilon  
}H_{\beta  \gamma  \delta  } \nabla^{\epsilon  }H_{\alpha  
}{}^{\gamma  \delta  } +c_{99} 
   \nabla_{\alpha  }T \nabla^{\alpha  }T \nabla_{\epsilon  
}H_{\beta  \gamma  \delta  } \nabla^{\epsilon  }H^{\beta  
\gamma  \delta  }  \nn\\&\!\!\!\!\!\!\!&+c_{100} 
   \nabla_{\alpha  }\Phi  \nabla^{\alpha  }\Phi  
\nabla_{\epsilon  }H_{\beta  \gamma  \delta  } 
\nabla^{\epsilon  }H^{\beta  \gamma  \delta  } + 
 c_{101} 
   H_{\alpha  }{}^{\beta  \gamma  }R_{\gamma  
\varepsilon  \delta  \epsilon  } \nabla^{\alpha  }\Phi  
\nabla^{\varepsilon  }H_{\beta  }{}^{\delta  \epsilon  } + 
 c_{102} 
   H_{\beta  \gamma  }{}^{\epsilon  } H^{\beta  \gamma  
\delta  } H_{\delta  }{}^{\varepsilon\mu} 
\nabla^{\alpha  }\Phi  \nabla_{\mu }H_{\alpha  
\epsilon  \varepsilon  }  \nn\\&\!\!\!\!\!\!\!&+c_{103} 
   H_{\alpha  }{}^{\beta  \gamma  } H_{\delta  \epsilon  }{}^{
\mu } H^{\delta  \epsilon  \varepsilon  } 
\nabla^{\alpha  }\Phi  \nabla_{\mu }H_{\beta  \gamma 
 \varepsilon  } +c_{104} 
   H_{\alpha  }{}^{\beta  \gamma  } H_{\beta  }{}^{\delta  
\epsilon  } H_{\delta  }{}^{\varepsilon\mu} 
\nabla^{\alpha  }\Phi  \nabla_{\mu }H_{\gamma  
\epsilon  \varepsilon  }  \nn\\&\!\!\!\!\!\!\!&+c_{105} 
   H_{\alpha  }{}^{\beta  \gamma  } H_{\beta  }{}^{\delta  
\epsilon  } H_{\gamma  }{}^{\varepsilon\mu} 
\nabla^{\alpha  }\Phi  \nabla_{\mu }H_{\delta  
\epsilon  \varepsilon  } \!+\!c_{106} 
   H_{\alpha  }{}^{\delta  \epsilon  } H^{\alpha  \beta  
\gamma  } \nabla_{\mu }H_{\delta  \epsilon  
\varepsilon  } \nabla^{\mu }H_{\beta  \gamma  
}{}^{\varepsilon  } \!+\!c_{107} 
   H_{\alpha  }{}^{\delta  \epsilon  } H^{\alpha  \beta  
\gamma  } \nabla_{\varepsilon  }H_{\gamma  \epsilon  
\mu } \nabla^{\mu }H_{\beta  \delta  
}{}^{\varepsilon  }  \nn\\&\!\!\!\!\!\!\!&+c_{108} 
   H_{\alpha  }{}^{\delta  \epsilon  } H^{\alpha  \beta  
\gamma  } \nabla_{\mu }H_{\gamma  \epsilon  
\varepsilon  } \nabla^{\mu }H_{\beta  \delta  
}{}^{\varepsilon  } +c_{109} 
   H_{\alpha  \beta  }{}^{\delta  } H^{\alpha  \beta  \gamma  
} \nabla_{\varepsilon  }H_{\delta  \epsilon\mu} 
\nabla^{\mu }H_{\gamma  }{}^{\epsilon  \varepsilon  } 
\!+\!c_{110} 
   H_{\alpha  \beta  }{}^{\delta  } H^{\alpha  \beta  \gamma  
} \nabla_{\mu }H_{\delta  \epsilon  \varepsilon  } 
\nabla^{\mu }H_{\gamma  }{}^{\epsilon  \varepsilon  
},\nn
\eeqa
where $c_1,\cdots, c_{110}$ are the coupling constants to be fixed by T-duality.

\subsection{ T-duality constraint on the minimal basis}

The above couplings should be fixed by the T-duality transformations in which the corrections to the Buscher rules include higher-derivative corrections to all base-space massless fields. The T-duality constraint at order \(\alpha'^2\) is given by \cite{Garousi:2019mca}:
\beqa
S^{(0,2)}_{(2)}(\psi_0) \sim S^{(2)}(\psi) - S^{(2)}(\psi_0) - S^{(0,2)}_{(2')}(\psi_0) - S^{(1,1)}(\psi_0)\,,\labell{T2}
\eeqa
where \(\psi\) collectively represents the massless fields in the base space, and \(\psi_0\) denotes their transformation under the Buscher rules \reef{flac1}. The corrections to the Buscher rules in \reef{flac1} at order \(\alpha'^n\), denoted by \(\psi^{(n)}\), are Taylor expanded in the above equation. The terms on the right-hand side of \reef{T2} are defined as follows. The last term corresponds to the Buscher transformation of the first-order perturbation of the reduction of the \(\alpha'\)-order action \reef{fourmax3}, which involves \(\psi^{(1)}\) as given in \reef{d1}. The third term is the Buscher transformation of the second-order perturbation of the reduction of the leading-order action \reef{S0}, which involves only terms quadratic in \(\psi^{(1)}\) from \reef{d1}. The first term is the reduction of \(\mathbf{S}^{(2)}\), while the second term is its transformation under the Buscher rules. All terms on the right-hand side depend only on the 110 coupling constants of \(\mathbf{S}^{(2)}\) and on some fixed numbers. The left-hand side of \reef{T2}, by contrast, is the Buscher transformation of the second-order perturbation of the reduction of the leading-order action \reef{S0}, which involves only \(\psi^{(2)}\). These perturbations include all contractions of the tachyon and massless base-space fields at order \(\alpha'^2\) with unknown coefficients, which therefore appear exclusively on the left-hand side of the equation.

One may consider only the independent corrections to the Buscher rules for which the corresponding \(S^{(0,2)}_{(2)}(\psi_0)\) is invariant under total derivative terms and the use of Bianchi identities. In that case, the above T-duality constraint fixes all coupling constants in the massless basis, as well as the parameters in the corrections to the Buscher rules. Alternatively, one may consider all possible corrections to the Buscher rules. Then the above T-duality constraint fixes the parameters in the massless basis, as well as some of the parameters in the corrections to the Buscher rules. However, there remain some parameters in the Buscher rules that are not fixed, and one is then allowed to set them all to zero. We use this latter approach. In the above equation, the symbol \(\sim\) means that the two sides are equal up to total derivative terms in the base space and the use of Bianchi identities. Importantly, the coupling-constant relations derived from the T-duality constraint \reef{T2} are independent of the base-space geometry \cite{Garousi:2019mca}. This allows us to considerably simplify the calculation by assuming a flat base space, which we adopt in the present work.

The above T-duality constraint produces 107 relations among the 110 parameters of the basis \reef{Lc}. One of them appears as the coefficient of \(\nabla^3 T \nabla^3 T\). Since the sphere-level S-matrix elements have tachyon poles that are produced by the standard kinetic term of the tachyon in the leading-order action, the coefficient of the above term, which modifies the tachyon propagator, must be zero. We therefore set this coefficient to zero. There are three T-duality multiplets. One of them includes six tachyons with coefficient \(c_{35}\), and another includes four tachyons with coefficient \(c_{56}\). The remaining T-dual multiplet has a fixed coefficient. They are
\beqa
\cL^{(2)}&\!\!\!=\!\!\!&
 - \frac{1}{192} H_{\alpha  }{}^{\delta  \epsilon  } 
H^{\alpha  \beta  \gamma  } H_{\beta  \delta  }{}^{\varepsilon  
} H_{\gamma  }{}^{\mu \zeta  } H_{\epsilon  
\mu }{}^{\eta  } H_{\varepsilon  \zeta  \eta  } + 
\frac{1}{480} H_{\alpha  \beta  }{}^{\delta  } H^{\alpha  
\beta  \gamma  } H_{\gamma  }{}^{\epsilon  \varepsilon  } 
H_{\delta  }{}^{\mu \zeta  } H_{\epsilon  \varepsilon  
}{}^{\eta  } H_{\mu \zeta  \eta  } \nn\\&\!\!\!\!\!\!\!&+ \frac{1}{320} 
H_{\alpha  \beta  }{}^{\delta  } H^{\alpha  \beta  \gamma  } 
H_{\gamma  }{}^{\epsilon  \varepsilon  } H_{\delta  \epsilon  
}{}^{\mu } H_{\varepsilon  }{}^{\zeta  \eta  } 
H_{\mu \zeta  \eta  } - \frac{1}{12} 
R_{\alpha  }{}^{\epsilon  }{}_{\gamma  }{}^{\varepsilon  
}R^{\alpha  \beta  \gamma  \delta  }R_{\beta 
 \epsilon  \delta  \varepsilon  } + \frac{1}{16} 
R_{\alpha  \beta  }{}^{\epsilon  \varepsilon  } 
R^{\alpha  \beta  \gamma  \delta  }R_{\gamma  
\delta  \epsilon  \varepsilon  } \nn\\&\!\!\!\!\!\!\!&+ \frac{1}{40} H_{\alpha  
}{}^{\delta  \epsilon  } H^{\alpha  \beta  \gamma  } H_{\beta  
}{}^{\varepsilon\mu} H_{\delta  \varepsilon  
}{}^{\zeta  }R_{\gamma  \epsilon\mu\zeta  
} - \frac{1}{8} H_{\alpha  }{}^{\delta  \epsilon  } H^{\alpha 
 \beta  \gamma  }R_{\beta  }{}^{\varepsilon  
}{}_{\delta  }{}^{\mu }R_{\gamma  \varepsilon 
 \epsilon\mu} + \frac{3}{320} H_{\alpha  \beta  
}{}^{\delta  } H^{\alpha  \beta  \gamma  } H_{\epsilon  
\varepsilon  }{}^{\zeta  } H^{\epsilon  \varepsilon \mu }R_{\gamma\mu\delta  \zeta  } \nn\\&\!\!\!\!\!\!\!&+ 
\frac{1}{8} H^{\alpha  \beta  \gamma  } H^{\delta  \epsilon  
\varepsilon  }R_{\alpha  \beta  \delta  
}{}^{\mu }R_{\gamma\mu\epsilon  
\varepsilon  } + \frac{1}{8} H_{\alpha  }{}^{\delta  \epsilon  
} H^{\alpha  \beta  \gamma  }R_{\beta  
}{}^{\varepsilon  }{}_{\delta  }{}^{\mu } 
R_{\gamma\mu\epsilon  \varepsilon  } - 
\frac{1}{8} H_{\alpha  }{}^{\delta  \epsilon  } H^{\alpha  
\beta  \gamma  }R_{\beta  }{}^{\varepsilon  
}{}_{\gamma  }{}^{\mu }R_{\delta  \varepsilon 
 \epsilon\mu} \nn\\&\!\!\!\!\!\!\!&- \frac{1}{16} H_{\alpha  \beta  
}{}^{\delta  } H^{\alpha  \beta  \gamma  }R_{\gamma  
}{}^{\epsilon  \varepsilon\mu}R_{\delta  
\varepsilon  \epsilon\mu} + \frac{3}{80} H_{\alpha 
 \beta  }{}^{\delta  } H^{\alpha  \beta  \gamma  } H_{\gamma  
}{}^{\epsilon  \varepsilon  } H_{\epsilon  }{}^{\mu 
\zeta  }R_{\delta\mu\varepsilon  \zeta  } 
+ \frac{1}{10} H_{\alpha  }{}^{\delta  \epsilon  } H^{\alpha  
\beta  \gamma  } H_{\beta  \delta  }{}^{\varepsilon  } 
H_{\gamma  }{}^{\mu \zeta  }R_{\epsilon  
\mu \varepsilon  \zeta  }\nn\\&\!\!\!\!\!\!\!& - \frac{1}{80} H_{\alpha  
\beta  }{}^{\delta  } H^{\alpha  \beta  \gamma  } H_{\gamma  
}{}^{\epsilon  \varepsilon  } H_{\delta  }{}^{\mu 
\zeta  }R_{\epsilon\mu\varepsilon  \zeta  
} + \frac{1}{768} H_{\beta  }{}^{\epsilon  \varepsilon  } 
H^{\beta  \gamma  \delta  } H_{\gamma  \epsilon  
}{}^{\mu } H_{\delta  \varepsilon\mu} 
\nabla_{\alpha  }T \nabla^{\alpha  }T \nn\\&\!\!\!\!\!\!\!&+ \frac{1}{32} 
R_{\beta  \gamma  \delta  \epsilon  }R^{\beta 
 \gamma  \delta  \epsilon  } \nabla_{\alpha  }T \nabla^{\alpha 
 }T\! - \!\frac{1}{16} H_{\beta  }{}^{\epsilon  \varepsilon  } H^{
\beta  \gamma  \delta  }R_{\gamma  \epsilon  \delta  
\varepsilon  } \nabla_{\alpha  }T \nabla^{\alpha  }T\!-\!
\frac{1}{64} H_{\alpha  }{}^{\gamma  \delta  } H_{\beta  
}{}^{\epsilon  \varepsilon  } H_{\gamma  \epsilon  
}{}^{\mu } H_{\delta  \varepsilon\mu} 
\nabla^{\alpha  }T \nabla^{\beta  }T \nn\\&\!\!\!\!\!\!\!&+ \frac{3}{40} H_{\alpha  
}{}^{\gamma  \delta  } H_{\beta  }{}^{\epsilon  \varepsilon  } 
R_{\gamma  \epsilon  \delta  \varepsilon  } 
\nabla^{\alpha  }T \nabla^{\beta  }T + \frac{3}{160} H_{\alpha 
 }{}^{\gamma  \delta  } H_{\beta  \gamma  }{}^{\epsilon  } 
H_{\delta  }{}^{\varepsilon\mu} H_{\epsilon  
\varepsilon\mu} \nabla^{\beta  }\nabla^{\alpha  
}\Phi \nn\\&\!\!\!\!\!\!\!& + \frac{3}{80} H_{\gamma  \delta  }{}^{\varepsilon  } 
H^{\gamma  \delta  \epsilon  }R_{\alpha  \epsilon  
\beta  \varepsilon  } \nabla^{\beta  }\nabla^{\alpha  }\Phi  + 
\frac{3}{20} H_{\alpha  }{}^{\gamma  \delta  } H_{\gamma  }{}^{
\epsilon  \varepsilon  }R_{\beta  \epsilon  \delta  
\varepsilon  } \nabla^{\beta  }\nabla^{\alpha  }\Phi  - 
\frac{3}{40} H_{\alpha  }{}^{\gamma  \delta  } H_{\beta  
}{}^{\epsilon  \varepsilon  }R_{\gamma  \epsilon  
\delta  \varepsilon  } \nabla^{\beta  }\nabla^{\alpha  }\Phi  
\nn\\&\!\!\!\!\!\!\!&+c_{35} 
   \nabla_{\alpha  }T \nabla^{\alpha  }T \nabla_{\beta  }T 
\nabla^{\beta  }T \nabla_{\gamma  }T \nabla^{\gamma  }T - 
\frac{1}{64} \nabla_{\alpha  }\Phi  \nabla^{\alpha  }T 
\nabla^{\beta  }T \nabla_{\gamma  }\nabla_{\beta  }T 
\nabla^{\gamma  }T \nn\\&\!\!\!\!\!\!\!&- \frac{1}{64} H_{\beta  }{}^{\delta  
\epsilon  } H_{\gamma  \delta  \epsilon  } \nabla^{\alpha  }T 
\nabla^{\beta  }\Phi  \nabla^{\gamma  }\nabla_{\alpha  }T +
  c_{56} 
   \nabla^{\alpha  }T \nabla^{\beta  }T \nabla_{\gamma  
}\nabla_{\beta  }T \nabla^{\gamma  }\nabla_{\alpha  }T\nn\\&\!\!\!\!\!\!\!& - 
\frac{1}{16} \nabla^{\beta  }\nabla^{\alpha  }T 
\nabla_{\gamma  }\nabla_{\beta  }\Phi  \nabla^{\gamma  
}\nabla_{\alpha  }T - \frac{3}{320} H_{\beta  }{}^{\delta  
\epsilon  } H_{\gamma  \delta  \epsilon  } \nabla_{\alpha  
}\Phi  \nabla^{\alpha  }T \nabla^{\gamma  }\nabla^{\beta  }T \nn\\&\!\!\!\!\!\!\!&+ 
\frac{1}{32} H_{\beta  }{}^{\delta  \epsilon  } \nabla^{\alpha 
 }T \nabla_{\gamma  }H_{\alpha  \delta  \epsilon  } 
\nabla^{\gamma  }\nabla^{\beta  }T - \frac{3}{80} H_{\alpha  
}{}^{\delta  \epsilon  } \nabla^{\alpha  }T \nabla_{\gamma  
}H_{\beta  \delta  \epsilon  } \nabla^{\gamma  }\nabla^{\beta  
}T \nn\\&\!\!\!\!\!\!\!&- \frac{1}{16} \nabla^{\alpha  }T \nabla_{\gamma  
}\nabla_{\beta  }\nabla_{\alpha  }T \nabla^{\gamma  
}\nabla^{\beta  }\Phi  + \frac{29}{80} H_{\alpha  \gamma  
}{}^{\epsilon  } H_{\beta  \delta  \epsilon  } \nabla^{\beta  }
\nabla^{\alpha  }T \nabla^{\delta  }\nabla^{\gamma  }T + 
\frac{3}{40}R_{\alpha  \gamma  \beta  \delta  } 
\nabla^{\beta  }\nabla^{\alpha  }T \nabla^{\delta  
}\nabla^{\gamma  }T \nn\\&\!\!\!\!\!\!\!&+ \frac{3}{160} H_{\alpha  \gamma  
}{}^{\epsilon  } H_{\beta  \delta  \epsilon  } \nabla^{\alpha  
}T \nabla^{\beta  }T \nabla^{\delta  }\nabla^{\gamma  }\Phi  + 
- \frac{31}{80}R_{\alpha  \beta  \gamma  \delta  } 
\nabla^{\alpha  }T \nabla^{\delta  }\nabla^{\gamma  
}\nabla^{\beta  }T\nn\\&\!\!\!\!\!\!\!& + \frac{33}{80} H_{\beta  }{}^{\delta  
\epsilon  } \nabla^{\alpha  }T \nabla^{\gamma  }\nabla^{\beta  
}T \nabla_{\epsilon  }H_{\alpha  \gamma  \delta  } + 
\frac{3}{20} \nabla^{\alpha  }T \nabla^{\beta  }T 
\nabla_{\delta  }H_{\beta  \gamma  \epsilon  } 
\nabla^{\epsilon  }H_{\alpha  }{}^{\gamma  \delta  }\nn\\&\!\!\!\!\!\!\!& - 
\frac{13}{160} \nabla^{\alpha  }T \nabla^{\beta  }T 
\nabla_{\epsilon  }H_{\beta  \gamma  \delta  } 
\nabla^{\epsilon  }H_{\alpha  }{}^{\gamma  \delta  } + 
\frac{3}{80} \nabla^{\beta  }\nabla^{\alpha  }\Phi  
\nabla_{\epsilon  }H_{\beta  \gamma  \delta  } 
\nabla^{\epsilon  }H_{\alpha  }{}^{\gamma  \delta  } \nn\\&\!\!\!\!\!\!\!&+ 
\frac{1}{192} \nabla_{\alpha  }T \nabla^{\alpha  }T 
\nabla_{\epsilon  }H_{\beta  \gamma  \delta  } 
\nabla^{\epsilon  }H^{\beta  \gamma  \delta  } + \frac{3}{160} 
H_{\beta  \gamma  }{}^{\epsilon  } H^{\beta  \gamma  \delta  } 
H_{\delta  }{}^{\varepsilon\mu} \nabla^{\alpha  
}\Phi  \nabla_{\mu }H_{\alpha  \epsilon  \varepsilon  
} \nn\\&\!\!\!\!\!\!\!&- \frac{3}{320} H_{\alpha  }{}^{\beta  \gamma  } H_{\delta  
\epsilon  }{}^{\mu } H^{\delta  \epsilon  \varepsilon  
} \nabla^{\alpha  }\Phi  \nabla_{\mu }H_{\beta  
\gamma  \varepsilon  } + \frac{1}{320} H_{\alpha  }{}^{\delta  
\epsilon  } H^{\alpha  \beta  \gamma  } \nabla_{\mu 
}H_{\delta  \epsilon  \varepsilon  } \nabla^{\mu 
}H_{\beta  \gamma  }{}^{\varepsilon  } \nn\\&\!\!\!\!\!\!\!&+ \frac{1}{80} H_{\alpha 
 }{}^{\delta  \epsilon  } H^{\alpha  \beta  \gamma  } 
\nabla_{\varepsilon  }H_{\gamma  \epsilon\mu} 
\nabla^{\mu }H_{\beta  \delta  }{}^{\varepsilon  } - 
\frac{1}{80} H_{\alpha  }{}^{\delta  \epsilon  } H^{\alpha  
\beta  \gamma  } \nabla_{\mu }H_{\gamma  \epsilon  
\varepsilon  } \nabla^{\mu }H_{\beta  \delta  
}{}^{\varepsilon  }\nn\\&\!\!\!\!\!\!\!& - \frac{1}{80} H_{\alpha  \beta  
}{}^{\delta  } H^{\alpha  \beta  \gamma  } \nabla_{\varepsilon  
}H_{\delta  \epsilon\mu} \nabla^{\mu }H_{
\gamma  }{}^{\epsilon  \varepsilon  } + \frac{1}{80} H_{\alpha  
\beta  }{}^{\delta  } H^{\alpha  \beta  \gamma  } 
\nabla_{\mu }H_{\delta  \epsilon  \varepsilon  } 
\nabla^{\mu }H_{\gamma  }{}^{\epsilon  \varepsilon  
}\,.\labell{SSS}
\eeqa
There are also lengthy expressions for the four-derivative corrections to the Buscher rules, which we refrain from writing out here.

Apart from the first term and the Riemann-cubed terms, which are invariant under field redefinitions and hence consistent with the S-matrix, all other couplings above with fixed coefficients are not consistent with the S-matrix, because in finding them the tachyon was treated as a massless field. To make them consistent with the S-matrix, we must use an appropriate field redefinition such that the new couplings become invariant under the restricted T-duality transformations. All such T-dual multiplets that include the above Riemann-cubed terms and are invariant under the restricted T-duality transformations are physically equivalent, because they should all be  consistent with the S-matrix elements. Since the tachyon is invariant under T-duality, such a T-dual multiplet may be constructed only from the massless fields.

To find this  T-dual multiplet for the massless fields, it is easier to first find the massless basis in which only Kalb–Ramond field redefinitions are used, and then determine its coupling constants by imposing invariance under the restricted T-duality transformations. Such a T-dual multiplet should be the same as the above T-dual multiplet with fixed coefficient, up to some field redefinition. As a result, the parameters in the newly constructed T-dual multiplet should be arbitrary. In the next section, we find such a T-dual multiplet.

\section{Massless sector at order $\alpha'^2$}

Using the xAct package \cite{Nutma:2013zea}, one finds that there are 705 couplings at order $\alpha'^2$ that involve covariant derivatives of the massless fields, as well as Riemann, Ricci, and scalar curvatures. Then, following \cite{Garousi:2019cdn}, and using total derivatives, Bianchi identities, and only Kalb–Ramond field redefinitions, one finds that there are 141 independent couplings. In a particular scheme, this massless basis is
\beqa
&\!\!\!\!\!\!\!&a_{1} 
   H_{\alpha  }{}^{\delta  \epsilon  } H^{\alpha  \beta  
\gamma  } H_{\beta  \delta  }{}^{\varepsilon  } H_{\gamma  }{}^{
\mu \zeta  } H_{\epsilon  \mu }{}^{\eta  } 
H_{\varepsilon  \zeta  \eta  } + a_{2} 
   H_{\alpha  \beta  }{}^{\delta  } H^{\alpha  \beta  \gamma  
} H_{\gamma  }{}^{\epsilon  \varepsilon  } H_{\delta  
}{}^{\mu \zeta  } H_{\epsilon  \mu }{}^{\eta 
 } H_{\varepsilon  \zeta  \eta  } \nn\\&\!\!\!\!\!\!\!&+ a_{3} 
   H_{\alpha  \beta  \gamma  } H^{\alpha  \beta  \gamma  } 
H_{\delta  }{}^{\mu \zeta  } H^{\delta  \epsilon  
\varepsilon  } H_{\epsilon  \mu }{}^{\eta  } 
H_{\varepsilon  \zeta  \eta  } + a_{4} 
   H_{\alpha  \beta  }{}^{\delta  } H^{\alpha  \beta  \gamma  
} H_{\gamma  }{}^{\epsilon  \varepsilon  } H_{\delta  
}{}^{\mu \zeta  } H_{\epsilon  \varepsilon  }{}^{\eta  
} H_{\mu \zeta  \eta  } \nn\\&\!\!\!\!\!\!\!& + a_{5} 
   H_{\alpha  \beta  }{}^{\delta  } H^{\alpha  \beta  \gamma  
} H_{\gamma  }{}^{\epsilon  \varepsilon  } H_{\delta  \epsilon  
}{}^{\mu } H_{\varepsilon  }{}^{\zeta  \eta  } 
H_{\mu \zeta  \eta  } + a_{6} 
   H_{\alpha  \beta  \gamma  } H^{\alpha  \beta  \gamma  } 
H_{\delta  \epsilon  }{}^{\mu } H^{\delta  \epsilon  
\varepsilon  } H_{\varepsilon  }{}^{\zeta  \eta  } 
H_{\mu \zeta  \eta  }  \nn\\&\!\!\!\!\!\!\!&+ a_{7} 
   H_{\alpha  \beta  \gamma  } H^{\alpha  \beta  \gamma  } 
H_{\delta  \epsilon  \varepsilon  } H^{\delta  \epsilon  
\varepsilon  } H_{\mu \zeta  \eta  } H^{\mu 
\zeta  \eta  } + a_{8} 
   H_{\alpha  }{}^{\gamma  \delta  } H_{\beta  }{}^{\epsilon  
\varepsilon  } H_{\gamma  \epsilon  }{}^{\mu } 
H_{\delta  \varepsilon  \mu } R^{\alpha  
\beta  } + a_{9} 
   H_{\alpha  }{}^{\gamma  \delta  } H_{\beta  }{}^{\epsilon  
\varepsilon  } H_{\gamma  \delta  }{}^{\mu } 
H_{\epsilon  \varepsilon  \mu } R^{\alpha  
\beta  } \nn\\&\!\!\!\!\!\!\!& + a_{10} 
   H_{\alpha  }{}^{\gamma  \delta  } H_{\beta  \gamma  
}{}^{\epsilon  } H_{\delta  }{}^{\varepsilon  \mu } H_{
\epsilon  \varepsilon  \mu } R^{\alpha  \beta 
 } + a_{11} 
   H_{\alpha  }{}^{\gamma  \delta  } H_{\beta  \gamma  \delta 
 } H_{\epsilon  \varepsilon  \mu } H^{\epsilon  
\varepsilon  \mu } R^{\alpha  \beta  } + 
   a_{12} 
   H_{\gamma  \delta  \epsilon  } H^{\gamma  \delta  \epsilon 
 } R_{\alpha  \beta  } R^{\alpha  \beta  } \nn\\&\!\!\!\!\!\!\!& +
    a_{13} 
   R_{\alpha  }{}^{\gamma  } R^{\alpha  \beta 
 } R_{\beta  \gamma  } + a_{14} 
   H_{\alpha  }{}^{\delta  \epsilon  } H^{\alpha  \beta  
\gamma  } H_{\beta  \delta  }{}^{\varepsilon  } H_{\gamma  
\epsilon  \varepsilon  } R + a_{15} 
   H_{\alpha  \beta  }{}^{\delta  } H^{\alpha  \beta  \gamma  
} H_{\gamma  }{}^{\epsilon  \varepsilon  } H_{\delta  \epsilon  
\varepsilon  } R  \nn\\&\!\!\!\!\!\!\!&+ a_{16} 
   H_{\alpha  \beta  \gamma  } H^{\alpha  \beta  \gamma  } 
H_{\delta  \epsilon  \varepsilon  } H^{\delta  \epsilon  
\varepsilon  } R + a_{17} 
   R_{\alpha  \beta  } R^{\alpha  \beta  }R + a_{18} 
   H_{\alpha  \beta  \gamma  } H^{\alpha  \beta  \gamma  } 
  R^2 + a_{19} R^3 + a_{20} 
   R R_{\alpha  \gamma  \beta  \delta  } 
R^{\alpha  \beta  \gamma  \delta  }  \nn\\&\!\!\!\!\!\!\!&+ a_{21} 
   H_{\alpha  }{}^{\delta  \epsilon  } H^{\alpha  \beta  
\gamma  } R R_{\beta  \delta  \gamma  \epsilon 
 } + a_{22} 
   R^{\alpha  \beta  } R_{\alpha  }{}^{\gamma 
 \delta  \epsilon  } R_{\beta  \delta  \gamma  
\epsilon  } + a_{23} 
   R_{\alpha  }{}^{\epsilon  }{}_{\gamma  
}{}^{\varepsilon  } R^{\alpha  \beta  \gamma  \delta  } 
R_{\beta  \varepsilon  \delta  \epsilon  } + 
   a_{24} 
   R_{\alpha  \beta  }{}^{\epsilon  \varepsilon  } 
R^{\alpha  \beta  \gamma  \delta  } R_{\gamma  
\epsilon  \delta  \varepsilon  } \nn\\&\!\!\!\!\!\!\!& + a_{25} 
   H_{\alpha  }{}^{\delta  \epsilon  } H^{\alpha  \beta  
\gamma  } H_{\beta  }{}^{\varepsilon  \mu } H_{\delta  
\varepsilon  }{}^{\zeta  } R_{\gamma  \epsilon  
\mu \zeta  } + a_{26} 
   H_{\alpha  \beta  }{}^{\delta  } H^{\alpha  \beta  \gamma  
} H_{\epsilon  \varepsilon  }{}^{\zeta  } H^{\epsilon  
\varepsilon  \mu } R_{\gamma  \mu 
\delta  \zeta  } + a_{27} 
   H_{\alpha  }{}^{\delta  \epsilon  } H^{\alpha  \beta  
\gamma  } R_{\beta  }{}^{\varepsilon  }{}_{\delta  }{}^{
\mu } R_{\gamma  \mu \epsilon  
\varepsilon  }  \nn\\&\!\!\!\!\!\!\!&+ a_{28} 
   H_{\alpha  }{}^{\delta  \epsilon  } H^{\alpha  \beta  
\gamma  } R_{\beta  }{}^{\varepsilon  }{}_{\gamma  }{}^{
\mu } R_{\delta  \varepsilon  \epsilon  
\mu } + a_{29} 
   H_{\alpha  \beta  }{}^{\delta  } H^{\alpha  \beta  \gamma  
} R_{\gamma  }{}^{\epsilon  \varepsilon  \mu } 
R_{\delta  \varepsilon  \epsilon  \mu } + 
   a_{30} 
   H_{\alpha  \beta  }{}^{\delta  } H^{\alpha  \beta  \gamma  
} H_{\gamma  }{}^{\epsilon  \varepsilon  } H_{\epsilon  
}{}^{\mu \zeta  } R_{\delta  \mu 
\varepsilon  \zeta  }  \nn\\&\!\!\!\!\!\!\!&+ a_{31} 
   H_{\alpha  \beta  \gamma  } H^{\alpha  \beta  \gamma  }R_{\delta  \varepsilon  \epsilon  \mu } 
R^{\delta  \epsilon  \varepsilon  \mu } + 
   a_{32} 
   H_{\alpha  }{}^{\delta  \epsilon  } H^{\alpha  \beta  
\gamma  } H_{\beta  \delta  }{}^{\varepsilon  } H_{\gamma  }{}^{
\mu \zeta  } R_{\epsilon  \mu 
\varepsilon  \zeta  } + a_{33} 
   H_{\alpha  \beta  }{}^{\delta  } H^{\alpha  \beta  \gamma  
} H_{\gamma  }{}^{\epsilon  \varepsilon  } H_{\delta  
}{}^{\mu \zeta  } R_{\epsilon  \mu 
\varepsilon  \zeta  }  \nn\\&\!\!\!\!\!\!\!&+\! a_{34} 
   H_{\alpha  \beta  \gamma  } H^{\alpha  \beta  \gamma  } 
H_{\delta  }{}^{\mu \zeta  } H^{\delta  \epsilon  
\varepsilon  } R_{\epsilon  \mu \varepsilon  
\zeta  } \!+ \!a_{35} 
   H_{\beta  }{}^{\epsilon  \varepsilon  } H^{\beta  \gamma  
\delta  } H_{\gamma  \epsilon  }{}^{\mu } H_{\delta  
\varepsilon  \mu } \nabla_{\alpha  }\nabla^{\alpha  }
\Phi  \!+ \!a_{36} 
   H_{\beta  \gamma  }{}^{\epsilon  } H^{\beta  \gamma  
\delta  } H_{\delta  }{}^{\varepsilon  \mu } 
H_{\epsilon  \varepsilon  \mu } \nabla_{\alpha  
}\nabla^{\alpha  }\Phi  \nn\\&\!\!\!\!\!\!\!& + a_{37} 
   H_{\beta  \gamma  \delta  } H^{\beta  \gamma  \delta  } 
H_{\epsilon  \varepsilon  \mu } H^{\epsilon  
\varepsilon  \mu } \nabla_{\alpha  }\nabla^{\alpha  }
\Phi  + a_{38} 
   H_{\beta  \gamma  \delta  } H^{\beta  \gamma  \delta  }R \nabla_{\alpha  }\nabla^{\alpha  }\Phi  + 
   a_{39} 
   R^2 \nabla_{\alpha  }\nabla^{\alpha  }\Phi  \nn\\&\!\!\!\!\!\!\!& + 
   a_{40} 
   R_{\beta  \delta  \gamma  \epsilon  } 
R^{\beta  \gamma  \delta  \epsilon  } \nabla_{\alpha  
}\nabla^{\alpha  }\Phi  + a_{41} 
   H_{\beta  \gamma  }{}^{\epsilon  } H^{\beta  \gamma  
\delta  } H_{\delta  }{}^{\varepsilon  \mu } 
H_{\epsilon  \varepsilon  \mu } \nabla_{\alpha  }\Phi 
 \nabla^{\alpha  }\Phi  + a_{42} 
   H_{\beta  \gamma  \delta  } H^{\beta  \gamma  \delta  } 
H_{\epsilon  \varepsilon  \mu } H^{\epsilon  
\varepsilon  \mu } \nabla_{\alpha  }\Phi  
\nabla^{\alpha  }\Phi  \nn\\&\!\!\!\!\!\!\!& + a_{43} 
   H_{\gamma  \delta  \epsilon  } H^{\gamma  \delta  \epsilon 
 } R^{\alpha  \beta  } \nabla_{\beta  }\nabla_{\alpha  
}\Phi  + a_{44} 
   R^{\alpha  \beta  } R \nabla_{\beta  
}\nabla_{\alpha  }\Phi  + a_{45} 
   H_{\gamma  \delta  \epsilon  } H^{\gamma  \delta  \epsilon 
 } \nabla_{\alpha  }\nabla^{\alpha  }\Phi  \nabla_{\beta  
}\nabla^{\beta  }\Phi  \nn\\&\!\!\!\!\!\!\!& + a_{46} 
   R \nabla_{\alpha  }\nabla^{\alpha  }\Phi  
\nabla_{\beta  }\nabla^{\beta  }\Phi  + a_{47} 
   H_{\gamma  \delta  \epsilon  } H^{\gamma  \delta  \epsilon 
 } \nabla_{\alpha  }\Phi  \nabla^{\alpha  }\Phi  
\nabla_{\beta  }\nabla^{\beta  }\Phi  + a_{48} 
   R \nabla_{\alpha  }\Phi  \nabla^{\alpha  }\Phi  
\nabla_{\beta  }\nabla^{\beta  }\Phi  \nn\\&\!\!\!\!\!\!\!& + a_{49} 
   \nabla_{\alpha  }R^{\alpha  \beta  } 
\nabla_{\beta  }\nabla_{\gamma  }\nabla^{\gamma  }\Phi  + 
   a_{50} 
   H^{\gamma  \delta  \epsilon  } R^{\alpha  \beta  } 
\nabla_{\beta  }\nabla_{\epsilon  }H_{\alpha  \gamma  \delta  
} + a_{51} 
   H_{\alpha  }{}^{\gamma  \delta  } R^{\alpha  \beta 
 } \nabla_{\beta  }\nabla_{\epsilon  }H_{\gamma  \delta  
}{}^{\epsilon  }  \nn\\&\!\!\!\!\!\!\!&+ a_{52} 
   H^{\alpha  \beta  \gamma  } R_{\gamma  \varepsilon 
 \delta  \epsilon  } \nabla_{\beta  }\nabla^{\varepsilon  
}H_{\alpha  }{}^{\delta  \epsilon  } + a_{53} 
   H_{\alpha  }{}^{\gamma  \delta  } H_{\beta  }{}^{\epsilon  
\varepsilon  } H_{\gamma  \epsilon  }{}^{\mu } 
H_{\delta  \varepsilon  \mu } \nabla^{\alpha  }\Phi  
\nabla^{\beta  }\Phi  + a_{54} 
   H_{\alpha  }{}^{\gamma  \delta  } H_{\beta  }{}^{\epsilon  
\varepsilon  } H_{\gamma  \delta  }{}^{\mu } 
H_{\epsilon  \varepsilon  \mu } \nabla^{\alpha  }\Phi 
 \nabla^{\beta  }\Phi  \nn\\&\!\!\!\!\!\!\!& + a_{55} 
   H_{\alpha  }{}^{\gamma  \delta  } H_{\beta  \gamma  
}{}^{\epsilon  } H_{\delta  }{}^{\varepsilon  \mu } H_{
\epsilon  \varepsilon  \mu } \nabla^{\alpha  }\Phi  
\nabla^{\beta  }\Phi  + a_{56} 
   H_{\alpha  }{}^{\gamma  \delta  } H_{\beta  \gamma  \delta 
 } H_{\epsilon  \varepsilon  \mu } H^{\epsilon  
\varepsilon  \mu } \nabla^{\alpha  }\Phi  
\nabla^{\beta  }\Phi  + a_{57} 
   H_{\alpha  }{}^{\gamma  \delta  } H_{\beta  \gamma  \delta 
 } R \nabla^{\alpha  }\Phi  \nabla^{\beta  }\Phi \nn\\&\!\!\!\!\!\!\!&  +
    a_{58} 
   H_{\alpha  }{}^{\gamma  \delta  } H_{\beta  }{}^{\epsilon  
\varepsilon  } R_{\gamma  \epsilon  \delta  \varepsilon 
 } \nabla^{\alpha  }\Phi  \nabla^{\beta  }\Phi  + 
   a_{59} 
   H_{\gamma  \delta  \epsilon  } H^{\gamma  \delta  \epsilon 
 } \nabla^{\alpha  }\Phi  \nabla_{\beta  }\nabla_{\alpha  
}\Phi  \nabla^{\beta  }\Phi  + a_{60} 
   R \nabla^{\alpha  }\Phi  \nabla_{\beta  }\nabla_{
\alpha  }\Phi  \nabla^{\beta  }\Phi   \nn\\&\!\!\!\!\!\!\!&+ a_{61} 
   \nabla_{\beta  }\nabla_{\gamma  }\nabla^{\gamma  }\Phi  
\nabla^{\beta  }\nabla_{\alpha  }\nabla^{\alpha  }\Phi  + 
   a_{62} 
   H_{\alpha  }{}^{\gamma  \delta  } H_{\beta  }{}^{\epsilon  
\varepsilon  } H_{\gamma  \epsilon  }{}^{\mu } 
H_{\delta  \varepsilon  \mu } \nabla^{\beta  
}\nabla^{\alpha  }\Phi  + a_{63} 
   H_{\alpha  }{}^{\gamma  \delta  } H_{\beta  }{}^{\epsilon  
\varepsilon  } H_{\gamma  \delta  }{}^{\mu } 
H_{\epsilon  \varepsilon  \mu } \nabla^{\beta  
}\nabla^{\alpha  }\Phi   \nn\\&\!\!\!\!\!\!\!&+ a_{64} 
   H_{\alpha  }{}^{\gamma  \delta  } H_{\beta  \gamma  
}{}^{\epsilon  } H_{\delta  }{}^{\varepsilon  \mu } H_{
\epsilon  \varepsilon  \mu } \nabla^{\beta  }\nabla^{
\alpha  }\Phi  + a_{65} 
   H_{\alpha  }{}^{\gamma  \delta  } H_{\beta  \gamma  \delta 
 } H_{\epsilon  \varepsilon  \mu } H^{\epsilon  
\varepsilon  \mu } \nabla^{\beta  }\nabla^{\alpha  
}\Phi  + a_{66} 
   H_{\alpha  }{}^{\gamma  \delta  } H_{\beta  \gamma  \delta 
 } R \nabla^{\beta  }\nabla^{\alpha  }\Phi  \nn\\&\!\!\!\!\!\!\!& + 
   a_{67} 
   R_{\alpha  }{}^{\gamma  \delta  \epsilon  } 
R_{\beta  \delta  \gamma  \epsilon  } \nabla^{\beta  }
\nabla^{\alpha  }\Phi  + a_{68} 
   H_{\gamma  \delta  \epsilon  } H^{\gamma  \delta  \epsilon 
 } \nabla_{\beta  }\nabla_{\alpha  }\Phi  \nabla^{\beta  
}\nabla^{\alpha  }\Phi  + a_{69} 
   R \nabla_{\beta  }\nabla_{\alpha  }\Phi  \nabla^{
\beta  }\nabla^{\alpha  }\Phi   \nn\\&\!\!\!\!\!\!\!&+ a_{70} 
   H^{\gamma  \delta  \epsilon  } \nabla_{\beta  
}\nabla_{\epsilon  }H_{\alpha  \gamma  \delta  } \nabla^{\beta 
 }\nabla^{\alpha  }\Phi  + a_{71} 
   \nabla_{\alpha  }R^{\alpha  \beta  } 
\nabla_{\gamma  }R_{\beta  }{}^{\gamma  } + 
   a_{72} 
   \nabla^{\alpha  }\Phi  \nabla^{\beta  }\nabla_{\alpha  
}\Phi  \nabla_{\gamma  }R_{\beta  }{}^{\gamma  }  \nn\\&\!\!\!\!\!\!\!&+ 
   a_{73} 
   R^{\beta  \gamma  } \nabla_{\alpha  }\Phi  
\nabla^{\alpha  }\Phi  \nabla_{\gamma  }\nabla_{\beta  }\Phi  
+ a_{74} 
   H_{\alpha  }{}^{\delta  \epsilon  } H_{\beta  \delta  
\epsilon  } R^{\alpha  \beta  } \nabla_{\gamma  
}\nabla^{\gamma  }\Phi  + a_{75} 
   R_{\alpha  \beta  } R^{\alpha  \beta  } 
\nabla_{\gamma  }\nabla^{\gamma  }\Phi  \nn\\&\!\!\!\!\!\!\!& + a_{76} 
   R^{\alpha  \beta  } \nabla_{\beta  
}\nabla_{\alpha  }\Phi  \nabla_{\gamma  }\nabla^{\gamma  
}\Phi  + a_{77} 
   \nabla_{\alpha  }\nabla^{\alpha  }\Phi  \nabla_{\beta  
}\nabla^{\beta  }\Phi  \nabla_{\gamma  }\nabla^{\gamma  }\Phi 
 + a_{78} 
   \nabla_{\alpha  }\Phi  \nabla^{\alpha  }\Phi  
\nabla_{\beta  }\nabla^{\beta  }\Phi  \nabla_{\gamma  
}\nabla^{\gamma  }\Phi  \nn\\&\!\!\!\!\!\!\!& + a_{79} 
   H_{\alpha  }{}^{\delta  \epsilon  } H_{\beta  \delta  
\epsilon  } \nabla^{\alpha  }\Phi  \nabla^{\beta  }\Phi  
\nabla_{\gamma  }\nabla^{\gamma  }\Phi  \!+\! a_{80} 
   \nabla_{\alpha  }\Phi  \nabla^{\alpha  }\Phi  
\nabla_{\beta  }\Phi  \nabla^{\beta  }\Phi  \nabla_{\gamma  }
\nabla^{\gamma  }\Phi  \!+\! a_{81} 
   \nabla^{\alpha  }\Phi  \nabla_{\beta  }\nabla_{\alpha  
}\Phi  \nabla^{\beta  }\Phi  \nabla_{\gamma  }\nabla^{\gamma  
}\Phi  \nn\\&\!\!\!\!\!\!\!& + a_{82} 
   \nabla^{\alpha  }\Phi  \nabla_{\beta  }\nabla^{\beta  
}\Phi  \nabla_{\gamma  }\nabla^{\gamma  }\nabla_{\alpha  
}\Phi  + a_{83} 
   \nabla_{\alpha  }R^{\alpha  \beta  } 
\nabla_{\gamma  }\nabla^{\gamma  }\nabla_{\beta  }\Phi  + 
   a_{84} 
   \nabla_{\alpha  }\nabla^{\beta  }\nabla^{\alpha  }\Phi  
\nabla_{\gamma  }\nabla^{\gamma  }\nabla_{\beta  }\Phi  \nn\\&\!\!\!\!\!\!\!& + 
   a_{85} 
   \nabla^{\beta  }\nabla_{\alpha  }\nabla^{\alpha  }\Phi  
\nabla_{\gamma  }\nabla^{\gamma  }\nabla_{\beta  }\Phi  + 
   a_{86} 
   H^{\alpha  \beta  \gamma  } R \nabla_{\gamma  
}\nabla_{\delta  }H_{\alpha  \beta  }{}^{\delta  } + 
   a_{87} 
   H_{\alpha  }{}^{\delta  \epsilon  } H^{\alpha  \beta  
\gamma  } H_{\beta  }{}^{\varepsilon  \mu } 
\nabla_{\gamma  }\nabla_{\mu }H_{\delta  \epsilon  
\varepsilon  } \nn\\&\!\!\!\!\!\!\!& + a_{88} 
   \nabla_{\beta  }R_{\alpha  \gamma  } 
\nabla^{\gamma  }R^{\alpha  \beta  } + a_{89}
    \nabla_{\beta  }\nabla_{\gamma  }\nabla_{\alpha  }\Phi  
\nabla^{\gamma  }R^{\alpha  \beta  } + a_{90}
    \nabla_{\gamma  }R_{\alpha  \beta  } 
\nabla^{\gamma  }R^{\alpha  \beta  } + a_{91}
    \nabla_{\gamma  }\nabla_{\beta  }\nabla_{\alpha  }\Phi  
\nabla^{\gamma  }R^{\alpha  \beta  }  \nn\\&\!\!\!\!\!\!\!&+ a_{92}
    H_{\beta  }{}^{\delta  \epsilon  } H_{\gamma  \delta  
\epsilon  } \nabla_{\alpha  }\Phi  \nabla^{\alpha  }\Phi  
\nabla^{\beta  }\Phi  \nabla^{\gamma  }\Phi  + a_{93}
    \nabla_{\alpha  }\Phi  \nabla^{\alpha  }\Phi  
\nabla^{\beta  }\Phi  \nabla_{\gamma  }\nabla_{\beta  }\Phi  
\nabla^{\gamma  }\Phi  + a_{94} 
   H_{\beta  }{}^{\delta  \epsilon  } H_{\gamma  \delta  
\epsilon  } R^{\alpha  \beta  } \nabla^{\gamma  
}\nabla_{\alpha  }\Phi   \nn\\&\!\!\!\!\!\!\!&+ a_{95} 
   H_{\beta  }{}^{\delta  \epsilon  } H_{\gamma  \delta  
\epsilon  } \nabla^{\alpha  }\Phi  \nabla^{\beta  }\Phi  
\nabla^{\gamma  }\nabla_{\alpha  }\Phi  + a_{96} 
   H_{\beta  }{}^{\delta  \epsilon  } H_{\gamma  \delta  
\epsilon  } \nabla^{\beta  }\nabla^{\alpha  }\Phi  
\nabla^{\gamma  }\nabla_{\alpha  }\Phi  + a_{97} 
   \nabla^{\alpha  }\Phi  \nabla^{\beta  }\Phi  
\nabla_{\gamma  }\nabla_{\beta  }\Phi  \nabla^{\gamma  
}\nabla_{\alpha  }\Phi  \nn\\&\!\!\!\!\!\!\!& + a_{98} 
   \nabla^{\beta  }\nabla^{\alpha  }\Phi  \nabla_{\gamma  
}\nabla_{\beta  }\Phi  \nabla^{\gamma  }\nabla_{\alpha  }\Phi 
 + a_{99} 
   H_{\beta  }{}^{\delta  \epsilon  } H_{\gamma  \delta  
\epsilon  } \nabla_{\alpha  }\nabla^{\alpha  }\Phi  
\nabla^{\gamma  }\nabla^{\beta  }\Phi  + a_{100} 
   H_{\beta  }{}^{\delta  \epsilon  } H_{\gamma  \delta  
\epsilon  } \nabla_{\alpha  }\Phi  \nabla^{\alpha  }\Phi  
\nabla^{\gamma  }\nabla^{\beta  }\Phi  \nn\\&\!\!\!\!\!\!\!& + a_{101} 
   H_{\beta  }{}^{\delta  \epsilon  } \nabla^{\alpha  }\Phi  
\nabla_{\gamma  }H_{\alpha  \delta  \epsilon  } \nabla^{\gamma 
 }\nabla^{\beta  }\Phi  + a_{102} 
   H_{\alpha  }{}^{\delta  \epsilon  } \nabla^{\alpha  }\Phi  
\nabla_{\gamma  }H_{\beta  \delta  \epsilon  } \nabla^{\gamma  
}\nabla^{\beta  }\Phi  + a_{103} 
   \nabla_{\alpha  }\nabla^{\alpha  }\Phi  \nabla_{\gamma  }
\nabla_{\beta  }\Phi  \nabla^{\gamma  }\nabla^{\beta  }\Phi  
 \nn\\&\!\!\!\!\!\!\!&+ a_{104} 
   \nabla_{\alpha  }\Phi  \nabla^{\alpha  }\Phi  
\nabla_{\gamma  }\nabla_{\beta  }\Phi  \nabla^{\gamma  
}\nabla^{\beta  }\Phi  + a_{105} 
   \nabla^{\alpha  }\Phi  \nabla_{\gamma  }\nabla_{\beta  
}\nabla_{\alpha  }\Phi  \nabla^{\gamma  }\nabla^{\beta  }\Phi 
 + a_{106} 
   \nabla_{\beta  }\nabla_{\gamma  }\nabla_{\alpha  }\Phi  
\nabla^{\gamma  }\nabla^{\beta  }\nabla^{\alpha  }\Phi  \nn\\&\!\!\!\!\!\!\!& + 
   a_{107} 
   \nabla_{\gamma  }\nabla_{\beta  }\nabla_{\alpha  }\Phi  
\nabla^{\gamma  }\nabla^{\beta  }\nabla^{\alpha  }\Phi  + 
   a_{108} 
   H^{\alpha  \beta  \gamma  } R \nabla_{\delta  
}\nabla^{\delta  }H_{\alpha  \beta  \gamma  } + 
   a_{109} 
   H_{\alpha  }{}^{\gamma  \delta  } R^{\alpha  \beta 
 } \nabla_{\delta  }\nabla_{\epsilon  }H_{\beta  \gamma  
}{}^{\epsilon  }  \nn\\&\!\!\!\!\!\!\!&+ a_{110} 
   H^{\alpha  \beta  \gamma  } R_{\beta  \epsilon  
\gamma  \varepsilon  } \nabla_{\delta  }\nabla^{\varepsilon  
}H_{\alpha  }{}^{\delta  \epsilon  } + a_{111} 
   H_{\alpha  \beta  }{}^{\delta  } H^{\alpha  \beta  \gamma  
} H^{\epsilon  \varepsilon  \mu } \nabla_{\delta  
}\nabla_{\mu }H_{\gamma  \epsilon  \varepsilon  } +
    a_{112} 
   H_{\alpha  \beta  }{}^{\delta  } H^{\alpha  \beta  \gamma  
} H_{\gamma  }{}^{\epsilon  \varepsilon  } \nabla_{\delta  
}\nabla_{\mu }H_{\epsilon  \varepsilon  
}{}^{\mu } \nn\\&\!\!\!\!\!\!\!& + a_{113} 
   H_{\alpha  \gamma  }{}^{\epsilon  } H_{\beta  \delta  
\epsilon  } R^{\alpha  \beta  } \nabla^{\delta  
}\nabla^{\gamma  }\Phi  + a_{114} 
   R^{\alpha  \beta  } R_{\alpha  \gamma  
\beta  \delta  } \nabla^{\delta  }\nabla^{\gamma  }\Phi  +
    a_{115} 
   H_{\alpha  \gamma  }{}^{\epsilon  } H_{\beta  \delta  
\epsilon  } \nabla^{\alpha  }\Phi  \nabla^{\beta  }\Phi  
\nabla^{\delta  }\nabla^{\gamma  }\Phi   \nn\\&\!\!\!\!\!\!\!&+ a_{116} 
   H_{\alpha  \gamma  }{}^{\epsilon  } H_{\beta  \delta  
\epsilon  } \nabla^{\beta  }\nabla^{\alpha  }\Phi  
\nabla^{\delta  }\nabla^{\gamma  }\Phi  + a_{117} 
   R_{\alpha  \gamma  \beta  \delta  } \nabla^{\beta 
 }\nabla^{\alpha  }\Phi  \nabla^{\delta  }\nabla^{\gamma  
}\Phi  + a_{118} 
   H^{\gamma  \delta  \epsilon  } \nabla^{\alpha  }\Phi  
\nabla_{\beta  }\nabla^{\beta  }\Phi  \nabla_{\epsilon  
}H_{\alpha  \gamma  \delta  }  \nn\\&\!\!\!\!\!\!\!&+ a_{119} 
   H_{\beta  }{}^{\delta  \epsilon  } \nabla^{\alpha  }\Phi  
\nabla^{\gamma  }\nabla^{\beta  }\Phi  \nabla_{\epsilon  
}H_{\alpha  \gamma  \delta  } + a_{120} 
   H^{\gamma  \delta  \epsilon  } \nabla^{\alpha  }\Phi  
\nabla^{\beta  }\nabla_{\alpha  }\Phi  \nabla_{\epsilon  
}H_{\beta  \gamma  \delta  } + a_{121} 
   H^{\gamma  \delta  \epsilon  } R^{\alpha  \beta  } 
\nabla_{\epsilon  }\nabla_{\beta  }H_{\alpha  \gamma  \delta  
}  \nn\\&\!\!\!\!\!\!\!&+ a_{122} 
   H^{\gamma  \delta  \epsilon  } \nabla^{\beta  
}\nabla^{\alpha  }\Phi  \nabla_{\epsilon  }\nabla_{\beta  }H_{
\alpha  \gamma  \delta  } + a_{123} 
   H_{\alpha  }{}^{\gamma  \delta  } R^{\alpha  \beta 
 } \nabla_{\epsilon  }\nabla_{\delta  }H_{\beta  \gamma  
}{}^{\epsilon  } + a_{124} 
   H_{\alpha  }{}^{\gamma  \delta  } \nabla^{\beta  }\nabla^{
\alpha  }\Phi  \nabla_{\epsilon  }\nabla_{\delta  }H_{\beta  
\gamma  }{}^{\epsilon  }  \nn\\&\!\!\!\!\!\!\!&+ a_{125} 
   H_{\alpha  }{}^{\gamma  \delta  } R^{\alpha  \beta 
 } \nabla_{\epsilon  }\nabla^{\epsilon  }H_{\beta  \gamma  
\delta  } + a_{126} 
   H^{\beta  \gamma  \delta  } \nabla_{\alpha  
}\nabla^{\alpha  }\Phi  \nabla_{\epsilon  }\nabla^{\epsilon  
}H_{\beta  \gamma  \delta  } + a_{127} 
   H_{\alpha  }{}^{\gamma  \delta  } \nabla^{\beta  }\nabla^{
\alpha  }\Phi  \nabla_{\epsilon  }\nabla^{\epsilon  }H_{\beta  
\gamma  \delta  }  \nn\\&\!\!\!\!\!\!\!&+ a_{128} 
   \nabla^{\beta  }\nabla^{\alpha  }\Phi  \nabla_{\delta  
}H_{\beta  \gamma  \epsilon  } \nabla^{\epsilon  }H_{\alpha  
}{}^{\gamma  \delta  } + a_{129} 
   \nabla^{\alpha  }\Phi  \nabla^{\beta  }\Phi  
\nabla_{\epsilon  }H_{\beta  \gamma  \delta  } 
\nabla^{\epsilon  }H_{\alpha  }{}^{\gamma  \delta  } + 
   a_{130} 
   \nabla^{\beta  }\nabla^{\alpha  }\Phi  \nabla_{\epsilon  
}H_{\beta  \gamma  \delta  } \nabla^{\epsilon  }H_{\alpha  
}{}^{\gamma  \delta  } \nn\\&\!\!\!\!\!\!\!& + a_{131} 
   \nabla_{\alpha  }\nabla^{\alpha  }\Phi  \nabla_{\epsilon  
}H_{\beta  \gamma  \delta  } \nabla^{\epsilon  }H^{\beta  
\gamma  \delta  } + a_{132} 
   \nabla_{\epsilon  }\nabla_{\delta  }H_{\alpha  \beta  
\gamma  } \nabla^{\epsilon  }\nabla^{\delta  }H^{\alpha  \beta 
 \gamma  } + a_{133} 
   H^{\alpha  \beta  \gamma  } R_{\beta  \delta  
\gamma  \epsilon  } \nabla_{\varepsilon  }\nabla^{\varepsilon  
}H_{\alpha  }{}^{\delta  \epsilon  }  \nn\\&\!\!\!\!\!\!\!&+ a_{134} 
   H_{\alpha  }{}^{\delta  \epsilon  } H^{\alpha  \beta  
\gamma  } H_{\beta  \delta  }{}^{\varepsilon  } 
\nabla_{\varepsilon  }\nabla_{\mu }H_{\gamma  
\epsilon  }{}^{\mu } + a_{135} 
   H^{\alpha  \beta  \gamma  } R_{\beta  \epsilon  
\gamma  \varepsilon  } \nabla^{\varepsilon  }\nabla_{\delta  
}H_{\alpha  }{}^{\delta  \epsilon  } + a_{136} 
   H^{\alpha  \beta  \gamma  } R_{\gamma  \epsilon  
\delta  \varepsilon  } \nabla^{\varepsilon  }\nabla^{\epsilon  
}H_{\alpha  \beta  }{}^{\delta  }  \nn\\&\!\!\!\!\!\!\!&+ a_{137} 
   H^{\alpha  \beta  \gamma  } R_{\gamma  \varepsilon 
 \delta  \epsilon  } \nabla^{\varepsilon  }\nabla^{\epsilon  
}H_{\alpha  \beta  }{}^{\delta  } + a_{138} 
   H_{\alpha  }{}^{\beta  \gamma  } H_{\beta  }{}^{\delta  
\epsilon  } H_{\gamma  }{}^{\varepsilon  \mu } 
\nabla^{\alpha  }\Phi  \nabla_{\mu }H_{\delta  
\epsilon  \varepsilon  } + a_{139} 
   H_{\alpha  }{}^{\delta  \epsilon  } H^{\alpha  \beta  
\gamma  } \nabla_{\varepsilon  }H_{\gamma  \epsilon  
\mu } \nabla^{\mu }H_{\beta  \delta  
}{}^{\varepsilon  }  \nn\\&\!\!\!\!\!\!\!&+ a_{140} 
   H_{\alpha  \beta  }{}^{\delta  } H^{\alpha  \beta  \gamma  
} \nabla_{\mu }H_{\delta  \epsilon  \varepsilon  } 
\nabla^{\mu }H_{\gamma  }{}^{\epsilon  \varepsilon  } 
+ a_{141} 
   H_{\alpha  \beta  \gamma  } H^{\alpha  \beta  \gamma  } 
\nabla_{\mu }H_{\delta  \epsilon  \varepsilon  } 
\nabla^{\mu }H^{\delta  \epsilon  \varepsilon  }\,.\labell{base}
\eeqa
where \(a_1, \ldots, a_{141}\) are 141 coupling constants that we are going to determine by the  T-duality constraint given in \reef{T2}, in which the corrections to the Buscher rules are restricted to the transformations in the second line of \reef{flac1}.

We find that the T-duality constraint \reef{T2} leaves 12 parameters in the massless basis \reef{base} arbitrary. We fix 9 of them by imposing the condition that the derivatives of the Ricci curvature and the third derivative of the dilaton vanish. We also set the other 3 parameters to zero. We find the following final 83 couplings:
\beqa
\cL_M^{(2)}&\!\!\!=\!\!\!&- \frac{1}{192} H_{\alpha }{}^{\delta \epsilon } H^{\alpha 
\beta \gamma } H_{\beta \delta }{}^{\varepsilon } H_{\gamma 
}{}^{\mu \zeta } H_{\epsilon \mu}{}^{\eta } 
H_{\varepsilon \zeta \eta } + \frac{1}{64} H_{\alpha \beta 
}{}^{\delta } H^{\alpha \beta \gamma } H_{\gamma }{}^{\epsilon 
\varepsilon } H_{\delta }{}^{\mu \zeta } H_{\epsilon 
\mu}{}^{\eta } H_{\varepsilon \zeta \eta }\nn\\&\!\!\!\!\!& -  
\frac{1}{2304} H_{\alpha \beta \gamma } H^{\alpha \beta 
\gamma } H_{\delta }{}^{\mu \zeta } H^{\delta 
\epsilon \varepsilon } H_{\epsilon \mu}{}^{\eta } 
H_{\varepsilon \zeta \eta } + \frac{1}{768} H_{\alpha \beta 
}{}^{\delta } H^{\alpha \beta \gamma } H_{\gamma }{}^{\epsilon 
\varepsilon } H_{\delta }{}^{\mu \zeta } H_{\epsilon 
\varepsilon }{}^{\eta } H_{\mu \zeta \eta }\nn\\&\!\!\!\!\!& + 
\frac{5}{256} H_{\alpha \beta }{}^{\delta } H^{\alpha \beta 
\gamma } H_{\gamma }{}^{\epsilon \varepsilon } H_{\delta 
\epsilon }{}^{\mu} H_{\varepsilon }{}^{\zeta \eta } 
H_{\mu \zeta \eta } -  \frac{1}{768} H_{\alpha \beta 
\gamma } H^{\alpha \beta \gamma } H_{\delta \epsilon 
}{}^{\mu} H^{\delta \epsilon \varepsilon } 
H_{\varepsilon }{}^{\zeta \eta } H_{\mu \zeta \eta } 
\nn\\&\!\!\!\!\!&+ \frac{1}{64} H_{\alpha }{}^{\gamma \delta } H_{\beta 
}{}^{\epsilon \varepsilon } H_{\gamma \epsilon }{}^{\varepsilon 
1} H_{\delta \varepsilon \mu} R^{\alpha \beta 
} -  \frac{1}{64} H_{\alpha }{}^{\gamma \delta } H_{\beta }{}^{
\epsilon \varepsilon } H_{\gamma \delta }{}^{\mu} 
H_{\epsilon \varepsilon \mu} R^{\alpha \beta 
} -  \frac{7}{64} H_{\alpha }{}^{\gamma \delta } H_{\beta 
\gamma }{}^{\epsilon } H_{\delta }{}^{\varepsilon \mu} 
H_{\epsilon \varepsilon \mu} R^{\alpha \beta 
}\nn\\&\!\!\!\!\!& + \frac{5}{768} H_{\alpha }{}^{\gamma \delta } H_{\beta 
\gamma \delta } H_{\epsilon \varepsilon \mu} 
H^{\epsilon \varepsilon \mu} R^{\alpha \beta 
} -  \frac{3}{16} H_{\beta }{}^{\delta \epsilon } H_{\gamma 
\delta \epsilon } R_{\alpha }{}^{\gamma } 
R^{\alpha \beta } -  \frac{3}{16} H_{\alpha \gamma 
}{}^{\epsilon } H_{\beta \delta \epsilon } R^{\alpha 
\beta } R^{\gamma \delta }\nn\\&\!\!\!\!\!& + \frac{1}{192} H_{\alpha 
}{}^{\delta \epsilon } H^{\alpha \beta \gamma } H_{\beta 
\delta }{}^{\varepsilon } H_{\gamma \epsilon \varepsilon } 
R + \frac{1}{64} H_{\alpha \beta }{}^{\delta } 
H^{\alpha \beta \gamma } H_{\gamma }{}^{\epsilon \varepsilon } 
H_{\delta \epsilon \varepsilon } R -  \frac{1}{8} 
H_{\alpha }{}^{\gamma \delta } H_{\beta \gamma \delta } 
R^{\alpha \beta } R \nn\\&\!\!\!\!\!&+ \frac{1}{4} H_{\gamma 
\delta }{}^{\varepsilon } H^{\gamma \delta \epsilon } 
R^{\alpha \beta } R_{\alpha \epsilon \beta 
\varepsilon } + \frac{1}{8} R R_{\alpha \beta 
\gamma \delta } R^{\alpha \beta \gamma \delta } -  
\frac{5}{8} R^{\alpha \beta } R_{\alpha 
}{}^{\gamma \delta \epsilon } R_{\beta \gamma \delta 
\epsilon } -  \frac{1}{8} H_{\alpha }{}^{\delta \epsilon } 
H^{\alpha \beta \gamma } R R_{\beta \delta 
\gamma \epsilon } \nn\\&\!\!\!\!\!&-  \frac{3}{4} H_{\alpha }{}^{\gamma \delta 
} H_{\gamma }{}^{\epsilon \varepsilon } R^{\alpha \beta 
} R_{\beta \epsilon \delta \varepsilon } -  
\frac{1}{12} R_{\alpha }{}^{\epsilon }{}_{\gamma 
}{}^{\varepsilon } R^{\alpha \beta \gamma \delta } 
R_{\beta \epsilon \delta \varepsilon } + \frac{1}{16} 
R_{\alpha \beta }{}^{\epsilon \varepsilon } R^{
\alpha \beta \gamma \delta } R_{\gamma \delta 
\epsilon \varepsilon } \nn\\&\!\!\!\!\!&+ \frac{3}{16} H_{\alpha }{}^{\gamma 
\delta } H_{\beta }{}^{\epsilon \varepsilon } R^{\alpha 
\beta } R_{\gamma \epsilon \delta \varepsilon } + 
\frac{35}{32} H^{\alpha \beta \gamma } H^{\delta \epsilon 
\varepsilon } R_{\alpha \beta \delta }{}^{\mu} R_{\gamma \epsilon \varepsilon \mu} -  
\frac{11}{16} H^{\alpha \beta \gamma } H^{\delta \epsilon 
\varepsilon } R_{\alpha \delta \beta }{}^{\mu} R_{\gamma \epsilon \varepsilon \mu} \nn\\&\!\!\!\!\!&+ 
\frac{1}{64} H_{\alpha }{}^{\delta \epsilon } H^{\alpha \beta 
\gamma } H_{\beta }{}^{\varepsilon \mu} H_{\delta 
\varepsilon }{}^{\zeta } R_{\gamma \epsilon \mu \zeta } -  \frac{35}{16} H_{\alpha }{}^{\delta \epsilon } H^{
\alpha \beta \gamma } R_{\beta }{}^{\varepsilon 
}{}_{\delta }{}^{\mu} R_{\gamma \varepsilon 
\epsilon \mu} -  \frac{5}{128} H_{\alpha \beta 
}{}^{\delta } H^{\alpha \beta \gamma } H_{\epsilon \varepsilon 
}{}^{\zeta } H^{\epsilon \varepsilon \mu} 
R_{\gamma \mu \delta \zeta }\nn\\&\!\!\!\!\!& + \frac{37}{16} 
H_{\alpha }{}^{\delta \epsilon } H^{\alpha \beta \gamma } 
R_{\beta }{}^{\varepsilon }{}_{\delta }{}^{\mu} R_{\gamma \mu \epsilon \varepsilon } + 
\frac{7}{8} H_{\alpha }{}^{\delta \epsilon } H^{\alpha \beta 
\gamma } R_{\beta }{}^{\varepsilon }{}_{\gamma 
}{}^{\mu} R_{\delta \varepsilon \epsilon 
\mu} + \frac{43}{32} H_{\alpha \beta }{}^{\delta } 
H^{\alpha \beta \gamma } R_{\gamma }{}^{\epsilon 
\varepsilon \mu} R_{\delta \varepsilon 
\epsilon \mu}\nn\\&\!\!\!\!\!& + \frac{35}{128} H_{\alpha \beta 
}{}^{\delta } H^{\alpha \beta \gamma } H_{\gamma }{}^{\epsilon 
\varepsilon } H_{\epsilon }{}^{\mu \zeta } R_{
\delta \mu \varepsilon \zeta } -  \frac{1}{96} 
H_{\alpha \beta \gamma } H^{\alpha \beta \gamma } 
R_{\delta \epsilon \varepsilon \mu} 
R^{\delta \epsilon \varepsilon \mu} -  
\frac{1}{16} H_{\alpha }{}^{\delta \epsilon } H^{\alpha \beta 
\gamma } H_{\beta \delta }{}^{\varepsilon } H_{\gamma 
}{}^{\mu \zeta } R_{\epsilon \mu 
\varepsilon \zeta }\nn\\&\!\!\!\!\!& -  \frac{7}{64} H_{\alpha \beta 
}{}^{\delta } H^{\alpha \beta \gamma } H_{\gamma }{}^{\epsilon 
\varepsilon } H_{\delta }{}^{\mu \zeta } 
R_{\epsilon \mu \varepsilon \zeta } + 
\frac{7}{384} H_{\alpha \beta \gamma } H^{\alpha \beta \gamma 
} H_{\delta }{}^{\mu \zeta } H^{\delta \epsilon 
\varepsilon } R_{\epsilon \mu \varepsilon 
\zeta }\nn\\&\!\!\!\!\!& + \frac{1}{96} H_{\beta }{}^{\epsilon \varepsilon } H^{
\beta \gamma \delta } H_{\gamma \epsilon }{}^{\mu} 
H_{\delta \varepsilon \mu} \nabla_{\alpha 
}\nabla^{\alpha }\Phi + \frac{1}{16} H_{\beta \gamma 
}{}^{\epsilon } H^{\beta \gamma \delta } H_{\delta 
}{}^{\varepsilon \mu} H_{\epsilon \varepsilon 
\mu} \nabla_{\alpha }\nabla^{\alpha }\Phi + 
\frac{1}{4} R_{\beta \gamma \delta \epsilon } 
R^{\beta \gamma \delta \epsilon } \nabla_{\alpha 
}\nabla^{\alpha }\Phi \nn\\&\!\!\!\!\!&-  \frac{1}{16} H_{\beta \gamma 
}{}^{\epsilon } H^{\beta \gamma \delta } H_{\delta 
}{}^{\varepsilon \mu} H_{\epsilon \varepsilon 
\mu} \nabla_{\alpha }\Phi \nabla^{\alpha }\Phi + 
\frac{1}{32} H_{\alpha }{}^{\gamma \delta } H_{\beta 
}{}^{\epsilon \varepsilon } H_{\gamma \epsilon }{}^{\mu} H_{\delta \varepsilon \mu} \nabla^{\beta }\nabla^{
\alpha }\Phi\nn\\&\!\!\!\!\!& -  \frac{1}{32} H_{\alpha }{}^{\gamma \delta } 
H_{\beta }{}^{\epsilon \varepsilon } H_{\gamma \delta 
}{}^{\mu} H_{\epsilon \varepsilon \mu} 
\nabla^{\beta }\nabla^{\alpha }\Phi -  \frac{7}{32} H_{\alpha 
}{}^{\gamma \delta } H_{\beta \gamma }{}^{\epsilon } H_{\delta 
}{}^{\varepsilon \mu} H_{\epsilon \varepsilon 
\mu} \nabla^{\beta }\nabla^{\alpha }\Phi\nn\\&\!\!\!\!\!& + 
\frac{5}{384} H_{\alpha }{}^{\gamma \delta } H_{\beta \gamma 
\delta } H_{\epsilon \varepsilon \mu} H^{\epsilon 
\varepsilon \mu} \nabla^{\beta }\nabla^{\alpha }\Phi 
-  \frac{1}{4} H_{\alpha }{}^{\gamma \delta } H_{\beta \gamma 
\delta } R \nabla^{\beta }\nabla^{\alpha }\Phi + 
\frac{1}{2} H_{\gamma \delta }{}^{\varepsilon } H^{\gamma 
\delta \epsilon } R_{\alpha \epsilon \beta \varepsilon 
} \nabla^{\beta }\nabla^{\alpha }\Phi \nn\\&\!\!\!\!\!&-  \frac{1}{2} 
R_{\alpha }{}^{\gamma \delta \epsilon } 
R_{\beta \delta \gamma \epsilon } \nabla^{\beta 
}\nabla^{\alpha }\Phi + H_{\alpha }{}^{\gamma \delta } 
H_{\gamma }{}^{\epsilon \varepsilon } R_{\beta \epsilon 
\delta \varepsilon } \nabla^{\beta }\nabla^{\alpha }\Phi -  
\frac{3}{8} H_{\alpha }{}^{\delta \epsilon } H_{\beta \delta 
\epsilon } R^{\alpha \beta } \nabla_{\gamma 
}\nabla^{\gamma }\Phi \nn\\&\!\!\!\!\!& -  \frac{1}{8} H_{\beta }{}^{\delta 
\epsilon } H_{\gamma \delta \epsilon } R^{\alpha \beta 
} \nabla^{\gamma }\nabla_{\alpha }\Phi + \frac{1}{2} H_{\beta 
}{}^{\delta \epsilon } H_{\gamma \delta \epsilon } 
\nabla^{\beta }\nabla^{\alpha }\Phi \nabla^{\gamma 
}\nabla_{\alpha }\Phi -  \frac{3}{4} H_{\beta }{}^{\delta 
\epsilon } H_{\gamma \delta \epsilon } \nabla_{\alpha 
}\nabla^{\alpha }\Phi \nabla^{\gamma }\nabla^{\beta }\Phi \nn\\&\!\!\!\!\!&+ 
\frac{9}{4} H_{\beta }{}^{\delta \epsilon } \nabla^{\alpha 
}\Phi \nabla_{\gamma }H_{\alpha \delta \epsilon } 
\nabla^{\gamma }\nabla^{\beta }\Phi + \frac{1}{8} H^{\alpha 
\beta \gamma } R \nabla_{\delta }\nabla_{\gamma 
}H_{\alpha \beta }{}^{\delta } -  \frac{1}{24} H^{\alpha \beta 
\gamma } R \nabla_{\delta }\nabla^{\delta }H_{\alpha 
\beta \gamma }\nn\\&\!\!\!\!\!& - 2 R_{\alpha \beta \gamma \delta } 
\nabla^{\alpha }\Phi \nabla^{\delta }R^{\beta \gamma 
} + \frac{1}{4} H_{\alpha \gamma }{}^{\epsilon } H_{\beta 
\delta \epsilon } R^{\alpha \beta } \nabla^{\delta 
}\nabla^{\gamma }\Phi + \frac{5}{4} H_{\alpha \gamma 
}{}^{\epsilon } H_{\beta \delta \epsilon } \nabla^{\beta 
}\nabla^{\alpha }\Phi \nabla^{\delta }\nabla^{\gamma }\Phi\nn\\&\!\!\!\!\!& - 
4 R_{\alpha \beta \gamma \delta } \nabla^{\alpha 
}\Phi \nabla^{\delta }\nabla^{\gamma }\nabla^{\beta }\Phi + 
\frac{1}{4} H^{\gamma \delta \epsilon } \nabla^{\alpha }\Phi 
\nabla_{\beta }\nabla^{\beta }\Phi \nabla_{\epsilon 
}H_{\alpha \gamma \delta } + 5 H_{\beta }{}^{\delta \epsilon } 
\nabla^{\alpha }\Phi \nabla^{\gamma }\nabla^{\beta }\Phi 
\nabla_{\epsilon }H_{\alpha \gamma \delta }\nn\\&\!\!\!\!\!& -  H^{\gamma 
\delta \epsilon } \nabla^{\alpha }\Phi \nabla^{\beta 
}\nabla_{\alpha }\Phi \nabla_{\epsilon }H_{\beta \gamma 
\delta } + \frac{3}{16} H^{\gamma \delta \epsilon } 
R^{\alpha \beta } \nabla_{\epsilon }\nabla_{\beta }H_{
\alpha \gamma \delta } + \frac{7}{8} H^{\gamma \delta 
\epsilon } \nabla^{\beta }\nabla^{\alpha }\Phi 
\nabla_{\epsilon }\nabla_{\beta }H_{\alpha \gamma \delta } \nn\\&\!\!\!\!\!&+ 
\frac{3}{16} H_{\alpha }{}^{\gamma \delta } R^{\alpha 
\beta } \nabla_{\epsilon }\nabla_{\beta }H_{\gamma \delta 
}{}^{\epsilon } + \frac{3}{8} H_{\alpha }{}^{\gamma \delta }
 R^{\alpha \beta } \nabla_{\epsilon }\nabla_{\delta 
}H_{\beta \gamma }{}^{\epsilon } -  \frac{1}{4} H_{\alpha }{}^{
\gamma \delta } \nabla^{\beta }\nabla^{\alpha }\Phi 
\nabla_{\epsilon }\nabla_{\delta }H_{\beta \gamma 
}{}^{\epsilon }\nn\\&\!\!\!\!\!& + \frac{1}{16} H_{\alpha }{}^{\gamma \delta } 
R^{\alpha \beta } \nabla_{\epsilon }\nabla^{\epsilon 
}H_{\beta \gamma \delta } -  \frac{1}{24} H^{\beta \gamma 
\delta } \nabla_{\alpha }\nabla^{\alpha }\Phi 
\nabla_{\epsilon }\nabla^{\epsilon }H_{\beta \gamma \delta } - 
 \frac{5}{8} H_{\alpha }{}^{\gamma \delta } \nabla^{\beta 
}\nabla^{\alpha }\Phi \nabla_{\epsilon }\nabla^{\epsilon 
}H_{\beta \gamma \delta }\nn\\&\!\!\!\!\!& -  \frac{3}{2} \nabla^{\beta 
}\nabla^{\alpha }\Phi \nabla_{\delta }H_{\beta \gamma 
\epsilon } \nabla^{\epsilon }H_{\alpha }{}^{\gamma \delta } -  
\frac{1}{12} \nabla_{\alpha }\nabla^{\alpha }\Phi 
\nabla_{\epsilon }H_{\beta \gamma \delta } \nabla^{\epsilon 
}H^{\beta \gamma \delta } + \frac{9}{16} H^{\alpha \beta 
\gamma } R_{\beta \delta \gamma \epsilon } 
\nabla_{\varepsilon }\nabla^{\varepsilon }H_{\alpha }{}^{\delta 
\epsilon } \nn\\&\!\!\!\!\!&-  \frac{35}{32} H^{\alpha \beta \gamma } 
R_{\gamma \varepsilon \delta \epsilon } 
\nabla^{\varepsilon }\nabla_{\beta }H_{\alpha }{}^{\delta 
\epsilon } + \frac{5}{8} H^{\alpha \beta \gamma } 
R_{\beta \epsilon \gamma \varepsilon } 
\nabla^{\varepsilon }\nabla_{\delta }H_{\alpha }{}^{\delta 
\epsilon } -  \frac{35}{32} H^{\alpha \beta \gamma } 
R_{\gamma \epsilon \delta \varepsilon } 
\nabla^{\varepsilon }\nabla^{\epsilon }H_{\alpha \beta 
}{}^{\delta } \nn\\&\!\!\!\!\!&-  \frac{1}{8} H^{\alpha \beta \gamma } 
R_{\gamma \varepsilon \delta \epsilon } 
\nabla^{\varepsilon }\nabla^{\epsilon }H_{\alpha \beta 
}{}^{\delta } + \frac{1}{128} H_{\alpha }{}^{\delta \epsilon } 
H^{\alpha \beta \gamma } H_{\beta }{}^{\varepsilon \mu} \nabla_{\mu}\nabla_{\gamma }H_{\delta \epsilon 
\varepsilon } -  \frac{1}{32} H_{\alpha \beta }{}^{\delta } H^{
\alpha \beta \gamma } H^{\epsilon \varepsilon \mu} 
\nabla_{\mu}\nabla_{\delta }H_{\gamma \epsilon 
\varepsilon } \nn\\&\!\!\!\!\!&-  \frac{1}{16} H_{\alpha }{}^{\delta \epsilon } 
H^{\alpha \beta \gamma } H_{\beta \delta }{}^{\varepsilon } 
\nabla_{\mu}\nabla_{\varepsilon }H_{\gamma \epsilon 
}{}^{\mu} + \frac{1}{64} H_{\alpha }{}^{\delta 
\epsilon } H^{\alpha \beta \gamma } \nabla_{\varepsilon 
}H_{\gamma \epsilon \mu} \nabla^{\varepsilon 
1}H_{\beta \delta }{}^{\varepsilon } \nn\\&\!\!\!\!\!&+ \frac{3}{128} H_{\alpha 
\beta }{}^{\delta } H^{\alpha \beta \gamma } 
\nabla_{\mu}H_{\delta \epsilon \varepsilon } \nabla^{
\mu}H_{\gamma }{}^{\epsilon \varepsilon } -  
\frac{1}{768} H_{\alpha \beta \gamma } H^{\alpha \beta \gamma 
} \nabla_{\mu}H_{\delta \epsilon \varepsilon } 
\nabla^{\mu}H^{\delta \epsilon \varepsilon }\,.\labell{S2}
\eeqa
There are also lengthy expressions for the four-derivative corrections to the Buscher rules in the second line of \reef{flac1}, which we do not write out because they are needed only for calculations at order \(\alpha'^3\), in which we are not interested in this paper.

It has been observed in \cite{Garousi:2026imu} that the contact terms between two tachyons and two massless fields at second order in the Mandelstam variables, which are produced by the expansion of the sphere-level S-matrix element of two tachyons and two massless NS-NS vertex operators, can be reproduced in field theory by the Feynman amplitude constructed from the tachyon couplings in the leading-order action and the massless couplings at order \(\alpha'\) in \reef{fourmax3}. In the next subsection, we examine the same calculation for the massless couplings at order \(\alpha'^2\) that we have found in \reef{S2}.

\subsection{Field theory amplitude }

To explicitly compute the Feynman amplitudes in field theory, one should first transform the string-frame couplings found from T-duality to the Einstein frame using the relation \(G_{\mu\nu} = e^{\gamma\Phi} g_{\mu\nu}\), where \(\gamma = 4/(D-2)\). In this frame, the leading-order action that includes the tachyon potential and two-derivative couplings takes the form
 \beqa
 \bS_{\rm leading} &\!\!\!\!=\!\!\!\!& \frac{1}{\kappa^2}\int d^{26}x \, \sqrt{-g} \left[ 2\left( R\! -\! \gamma \partial_\mu \Phi \partial^\mu \Phi\!-\!\frac{1}{12}e^{-2\gamma \Phi}H^2 \right) \!-\! \frac{1}{2}\partial_\mu T \partial^\mu T \!-\!2 e^{\gamma\Phi} V(T) \right] . \labell{Sbulkbrane1}
 \eeqa
Similarly, one can transform the string-frame couplings in \reef{S2} to the Einstein frame,
\beqa
{\bf S}_{\mathrm{M}}^{(2)}&\!\!\!=\!\!\!&\frac{2}{\kappa^2}\int d^{26}x \sqrt{-g}\Big[  -  
\frac{e^{-3\gamma\Phi}}{12} R_{\alpha }{}^{\epsilon }{}_{\gamma 
}{}^{\varepsilon } R^{\alpha \beta \gamma \delta } 
R_{\beta \epsilon \delta \varepsilon } + \frac{e^{-3\gamma\Phi}}{16} 
R_{\alpha \beta }{}^{\epsilon \varepsilon } R^{
\alpha \beta \gamma \delta } R_{\gamma \delta 
\epsilon \varepsilon } +\cdots\Big],\labell{six}
\eeqa
where the dots represent many other terms that we do not write explicitly here.
We also need standard kinetic terms for the fluctuations, so we normalize the fields as
 \beqa
 g_{\mu\nu} = \eta_{\mu\nu} + \kappa h_{\mu\nu}, \qquad \Phi = \frac{\kappa}{2\sqrt{\gamma}} \phi, \qquad T = \kappa \tau, \qquad B_{\mu\nu}=\kappa b_{\mu\nu}\,.
 \eeqa

We are interested only in the term in the amplitude that has the factor \(\Tr(\epsilon_3 \cdot \epsilon_4)\), where \(\epsilon_3, \epsilon_4\) are the polarizations of the two gravitons. This appears in the \(s\)-channel amplitude, where the Mandelstam variables are
\beqa
s=-(p_1+p_2)^2,\qquad t=-(p_1+p_3)^2,\qquad u=-(p_1+p_4)^2 \,,
\eeqa
in the convension that $\alpha'=2$. 
Using the on-shell relations \(p_1^2 = p_2^2 = 2\) and \(p_3^2 = p_4^2 = 0\), one finds that the Mandelstam variables satisfy the following on-shell relation:
\beqa
s+t+u=-4\,.\labell{stu2T2G}
\eeqa
 This contribution in field theory up to order \(\alpha'^2\), i.e., from \(\!\!\bS = \bS_{\rm leading} + \alpha' \bS^{(1)} + \alpha'^2 \bS^{(2)}\), for two gravitons  is given by
\beqa
 A'_s &=& \tilde{V}_{\tau_1\tau_2\phi} \tilde{G}_\phi \tilde{V}_{\phi h_3h_4} + (\tilde{V}_{\tau_1\tau_2h})^{\mu\nu} (\tilde{G}_h)_{\mu\nu\alpha\beta} (\tilde{V}_{h h_3h_4})^{\alpha\beta}\,, \labell{A's1}
 \eeqa
where the propagators and the first vertex in each term are extracted from the leading-order action \reef{Sbulkbrane1}, 
\beqa
 \tilde{G}_\phi &=& -\frac{i}{k^2}, \qquad (\tilde{G}_h)_{\alpha\beta,\mu\nu} = -\frac{i}{2k^2} \left( \eta_{\alpha\mu}\eta_{\beta\nu} + \eta_{\alpha\nu}\eta_{\beta\mu} - \frac{2}{D-2}\eta_{\alpha\beta}\eta_{\mu\nu} \right), \nn\\
 \tilde{V}_{\tau_1\tau_2\phi } &=& i\kappa\sqrt{\gamma}, \qquad (\tilde{V}_{ \tau_1\tau_2 h})^{\mu\nu} = -\frac{i\kappa}{2} \left[ p_1^\mu p_2^\nu + p_2^\mu p_1^\nu - \eta^{\mu\nu} (p_1\cdot p_2 + 2) \right]\,.\labell{pro}
 \eeqa
The second vertices can be  extracted from the six-derivative action \reef{six}.  Substituting these expressions into \reef{A's1}, one finds the first and second terms on the right-hand side of \reef{A's1} to be
\beqa
 \tilde{V}_{\tau_1\tau_2\phi} \tilde{G}_\phi \tilde{V}_{\phi h_3h_4}&=&-\frac{i\kappa^2}{2(D-2)}s^2\Tr(\epsilon_3\cdot\epsilon_4)\,,\nn\\
  (\tilde{V}_{\tau_1\tau_2h})^{\mu\nu} (\tilde{G}_h)_{\mu\nu\alpha\beta} (\tilde{V}_{h h_3h_4})^{\alpha\beta}&=&-\frac{i\kappa^2}{16(D-2)}s\Big[-3t'u'(D-2)-8s\Big]\Tr(\epsilon_3\cdot\epsilon_4)\,,\nn
 \eeqa
where we have used  the on-shell relations \(p_1\cdot p_3 = p_2\cdot p_4 = -1 - t/2\), \(p_1\cdot p_4 = p_2\cdot p_3 = -1 - u/2\), \(p_1\cdot p_2 = -2 - s/2\), and \(p_3\cdot p_4 = -s/2\). The new Mandelstam variables $t',u',$ are defined as 
\beqa
t'=t+2=-2p_1\cdot p_3,\,\,\,u'=u+2=-2p_1\cdot p_4\,.
\eeqa
Adding the two contributions, one finds the amplitude \reef{A's1} becomes
\beqa
  A'^{(3)}_s&=&\frac{3i\kappa^2}{16}\Tr(\epsilon_3\cdot\epsilon_4^T)st'u'\,,\labell{A'3}
 \eeqa
which is at the third order of old and new Mandelstam variables. 
Note that the new Mandelstam variables \(t', u'\) would be the same as the Mandelstam variables \(t, u\) if the tachyon were a massless field.
 It is important to note that the above on-shell amplitude can not be written in terms of third order of only old Mandelstam variables $s,t,u$. We will see that the string theory amplitude also has expansion in terms of both old and new Mandelstam variables.

The field theory amplitude   for  two tachyons and  two Kalb-Ramond states   is given by
\beqa
 A'_s &=& \tilde{V}_{\tau_1\tau_2\phi} \tilde{G}_\phi \tilde{V}_{\phi b_3b_4} + (\tilde{V}_{\tau_1\tau_2h})^{\mu\nu} (\tilde{G}_h)_{\mu\nu\alpha\beta} (\tilde{V}_{h b_3b_4})^{\alpha\beta}\,, \labell{A's2}
 \eeqa
where the  propagators and the first vertex in each term are given in \reef{pro}, and the second vertices can be  extracted from the six-derivative action \reef{six}. In this case also we find  
\beqa
 \tilde{V}_{\tau_1\tau_2\phi} \tilde{G}_\phi \tilde{V}_{\phi b_3b_4}&=&-\frac{i\kappa^2}{2(D-2)}s^2\Tr(\epsilon_3\cdot\epsilon_4^T)\,,\nn\\
  (\tilde{V}_{\tau_1\tau_2h})^{\mu\nu} (\tilde{G}_h)_{\mu\nu\alpha\beta} (\tilde{V}_{h b_3b_4})^{\alpha\beta}&=&-\frac{i\kappa^2}{16(D-2)}s\Big[-3t'u'(D-2)-8s\Big]\Tr(\epsilon_3\cdot\epsilon_4^T)\,.\nn
 \eeqa
 Substituting them into \reef{A's2}, one finds the same result as in \reef{A'3}. In the next subsection we compare the field theory amplitude \reef{A'3} with the corresponding sphere-level S-matrix element in string theory.

\subsection{String theory amplitude }

The sphere-level S-matrix element of two tachyons and two massless NS-NS vertex operators contains some terms in which momenta are contracted with the NS-NS polarization tensors. For the term in which the polarizations of the NS-NS states contract with each other, the amplitude can be written as \cite{Garousi:2026imu}
\beqa
A&=&\frac{i\kappa^2}{2}\Tr(\epsilon_3\cdot\epsilon_4^T)\frac{\Gamma(-u/2)\Gamma(-1-s/2)\Gamma(-t/2)}{\Gamma(1+u/2)\Gamma(2+s/2)\Gamma(1+t/2)}\,,\labell{Amp}
\eeqa
which is symmetric under the exchange of \(t\) and \(u\). 
The expansion of the string amplitude \reef{Amp} should be consistent with the  relation \reef{stu2T2G} and should be reproduced by the leading term of the effective action \reef{Sbulkbrane1}. This forces us to rewrite the amplitude \reef{Amp} as \cite{Garousi:2026imu}
\beqa
A&=&-\frac{i\kappa^2}{2}\Tr(\epsilon_3\cdot\epsilon_4^T)\frac{1}{(1+s/2)^2}\frac{\Gamma(1-u'/2)\Gamma(-s/2)\Gamma(1-t'/2)}{\Gamma(u'/2)\Gamma(1+s/2)\Gamma(t'/2)}\,.\labell{ATG1}
\eeqa
Now the old and new Mandelstam variables satisfy the on-shell relation
\beqa
s+t'+u'=0\,.\labell{st'u'2T2G}
\eeqa
The amplitude contains tachyon and massless poles in the \(s\)-channel. There are no tachyon or massless poles in the \(t'\)-channel or the \(u'\)-channel. There is an infinite tower of massive poles in all channels. The expansion consistent with the on-shell relation \reef{st'u'2T2G} is the expansion in the limit \(s, t', u' \to 0\). This expands the tachyon pole in the \(s\)-channel and all massive poles in the \(s\)-, \(t'\)-, and \(u'\)-channels. The expansion of the tachyon pole is
\beqa
\frac{1}{(1+s/2)^2}&=&1-s+\frac{3}{4}s^2-\frac{1}{2}s^3+\cdots\,,\labell{exT1}
\eeqa
and the expansion of the massive poles is
\beqa
\frac{\Gamma(1-t'/2)\Gamma(-s/2)\Gamma(1-u'/2)}{\Gamma(t'/2)\Gamma(1+s/2)\Gamma(u'/2)}&=&-\frac{u't'}{2s}-\frac{(u't')^2}{8}\z(3)+\cdots\,.\labell{exG1}
\eeqa
Using the above expansion, the amplitude \reef{ATG1} produces the following expansion:
\beqa
A&=&\frac{i\kappa^2}{4}\Tr(\epsilon_3\cdot\epsilon_4^T)\Big[\frac{u't'}{s}-u't'+\frac{3}{4}su't'-\frac{1}{2}s^2u't'+\frac{1}{4}(u't')^2\z(3)+\cdots\Big]\,.\labell{AA}
\eeqa
It is important to note that the second, third, and fourth terms above result from expanding the tachyon pole in the amplitude \reef{Amp}. The first term is exactly reproduced by the leading-order effective action \reef{Sbulkbrane1}. It has been shown in \cite{Garousi:2026imu} that the leading-order effective action \reef{Sbulkbrane1} and the four-derivative effective action in \reef{fourmax3} exactly reproduce the second term in the above expansion. On the other hand, the third term above is also exactly the same as the field-theory amplitude \reef{A'3} that we have found from the leading-order effective action and the six-derivative effective action \reef{S2}. The fourth and fifth terms in the above expansion should be reproduced by the eight-derivative effective action, in which we are not interested in this paper.

Note also that the term with coefficient \(\zeta(3)\) appears in the expansion of the amplitude of two tachyons and two massless NS-NS vertex operators in type 0 theory \cite{Garousi:2003db}, in which the new Mandelstam variables are \(t' = t+1\) and \(u' = u+1\). Hence, this term should be reproduced by eight-derivative couplings in type 0 theory as well. In type 0 theory, however, there are no terms like the second, third, and fourth terms in the expansion \reef{AA}. It would be interesting to find the eight-derivative couplings in type 0 theory and reproduce the above term with coefficient \(\zeta(3)\).

\section{Tachyon sector at order $\alpha'^2$}

Having found the massless sector at order \(\alpha'^2\) that is consistent with T-duality and the string theory S-matrix element, we now consider the couplings in the tachyon sector. These couplings are found by constraining the minimal basis with T-duality in \reef{SSS}. They are
\beqa
\cL_T^{(2)}&=&c_{35} 
   \nabla_{\alpha  }T \nabla^{\alpha  }T \nabla_{\beta  }T 
\nabla^{\beta  }T \nabla_{\gamma  }T \nabla^{\gamma  }T  +
  c_{56} 
   \nabla^{\alpha  }T \nabla^{\beta  }T \nabla_{\gamma  
}\nabla_{\beta  }T \nabla^{\gamma  }\nabla_{\alpha  }T\,,\labell{TTTT}
\eeqa
where the coefficients \(c_{35}\) and \(c_{56}\) remain unfixed by T-duality. The above two couplings are invariant under the Buscher rules \reef{flac1}. Hence, we do not need to use any field redefinition to make them invariant under the restricted T-duality transformations. The coefficient \(c_{35}\) should be fixed by the expansion of the sphere-level S-matrix element of six tachyon vertex operators. Such an expansion should fix this coefficient as well as the coefficient of \(T^6\) in the tachyon potential. We leave this calculation for future work.

To find the coefficient \(c_{56}\), one must compare the S-matrix element produced by the above terms with the corresponding string theory S-matrix element. This coupling produces the following contact terms of four tachyon states in momentum space:
\beqa
A'^{(3)}_c&=&4i\kappa^2c_{56}\Big[8(4+t)^2+s^2(8+3t)+s(64+32t+3t^2)\Big]\,,\labell{A'tttt1}
\eeqa
where we have used conservation of momentum to write \(p_4\) in terms of \(p_1, p_2, p_3\). Then we use the on-shell relations \(p_1^2 = p_2^2 = p_3^2 = 2\), \(p_1 \cdot p_2 = -2 - s/2\), \(p_1 \cdot p_3 = -2 - u/2\), \(p_2 \cdot p_3 = -2 - t/2\), and the on-shell relation
\beqa
s+t+u=-8\,.\labell{stu4T}
\eeqa
One finds that the amplitude \reef{A'tttt1} can be written in terms of the old and new Mandelstam variables as
\beqa
A'^{(3)}_c&=&-4
i\kappa^2c_{56}(su't'+us't'+t u's')\,.\labell{A'tttt}
\eeqa
The new  Mandelstam variables $s',\,t', \, u'$ are  defined as
 \beqa
  s'\,=\,s+4=-2p_1\cdot p_2\,, \,\,\,t'\,=\,t+4=-2p_1\cdot p_4\,, \,\,\,u'\,=\,u+4=-2p_1\cdot p_3\,.
 \eeqa
 These variables appear in the expansion of string theory S-matrix element.

Now the string theory sphere-level S-matrix element of four tachyon vertex operator is given by the   well-known Virasoro–Shapiro amplitude \cite{Virasoro:1969me,Shapiro:1969km}
\beqa
A&=&\frac{3i\kappa^2}{2}\frac{\Gamma(-1-s/2)\Gamma(-1-t/2)\Gamma(-1-u/2)}{\Gamma(2+s/2)\Gamma(2+t/2)\Gamma(2+u/2)}\,.
\eeqa
In order for the expansion of this amplitude to reproduce the massless poles of field theory produced by the standard kinetic term of the tachyon in the leading-order effective action \reef{Sbulkbrane1}, one should rewrite the amplitude as  \cite{Garousi:2026imu}
 \beqa
A&=&-\frac{i\kappa^2}{2}\Big[\frac{1}{(1+s/2)^2}\frac{\Gamma(1-t'/2)\Gamma(-s/2)\Gamma(1-u'/2)}{\Gamma(t'/2)\Gamma(1+s/2)\Gamma(u'/2)}\nn\\&&\qquad\quad+\frac{1}{(1+t/2)^2}\frac{\Gamma(1-s'/2)\Gamma(-t/2)\Gamma(1-u'/2)}{\Gamma(s'/2)\Gamma(1+t/2)\Gamma(u'/2)}\nn\\&&\qquad\quad+\frac{1}{(1+u/2)^2}\frac{\Gamma(1-s'/2)\Gamma(-u/2)\Gamma(1-t'/2)}{\Gamma(s'/2)\Gamma(1+u/2)\Gamma(t'/2)}\Big]\,.\labell{Astring}
\eeqa
The first, second, and third lines in \reef{Astring} contain tachyon and massless poles in the \(s\)-, \(t\)-, and \(u\)-channels, respectively. The expansion is then performed in the limit \(s, t, u, s', t', u' \to 0\), keeping the massless poles and expanding the tachyon poles. Using the expansions in \reef{exT1} and \reef{exG1} for all three channels, one finds
\beqa
A&=&\frac{i\kappa^2}{4}\Big[\frac{u't'}{s}+\frac{u's'}{t}+\frac{t's'}{u}-(u't'+u's'+t's')+\frac{3}{4}(su't'+tu's'+us't')\nn\\
&&\qquad-\frac{1}{2}(s^2u't'+t^2u's'+u^2s't')+\frac{1}{4}[(u't')^2+(u's')^2+(s't')^2]\z(3)+\cdots\Big].\labell{Expand}
\eeqa
It has been shown in  \cite{Garousi:2026imu} that the massless poles are exactly the same as the field-theory massless poles. It has also been shown that the terms at second order in the Mandelstam variables are reproduced by the coupling in the tachyon sector of \reef{fourmax3} and by the quartic term in the tachyon potential \reef{VT1}. One can easily observe that the terms at third order in the Mandelstam variables in the above expansion are the same as the field-theory amplitude \reef{A'tttt}. This fixes the coefficient to \(c_{56} = -3/64\). Hence, the tachyon sector at order \(\alpha'^2\) becomes
\beqa
\cL_T^{(2)}&=&c_{35} \nabla_{\alpha  }T \nabla^{\alpha  }T 
\nabla_{\beta  }T \nabla^{\beta  }T \nabla_{\gamma  }T 
\nabla^{\gamma  }T - 
 \frac{3}{64} \nabla^{\alpha  }T \nabla^{\beta  }T 
\nabla_{\gamma  }\nabla_{\beta  }T \nabla^{\gamma  
}\nabla_{\alpha  }T\,.\labell{4T}
\eeqa
The coefficient \(c_{35}\) and the coefficient of \(T^6\) in the tachyon potential can be found by analyzing the S-matrix element of six tachyon vertex operators.
Moreover, the terms at fourth order in the Mandelstam variables in the expansion \reef{Expand} should be reproduced by the eight-derivative effective action, in which we are not interested in this paper.

It is important to note also that the term with coefficient \(\zeta(3)\) appears in the expansion of the amplitude of four tachyon vertex operators in type 0 theory \cite{Garousi:2003db}, in which the new Mandelstam variables are \(t' = t+2\), \(u' = u+2\), and \(s' = s+2\). Hence, this term should be reproduced by eight-derivative couplings in type 0 theory as well. In type 0 theory, however, the four tachyon amplitude has no tachyon poles, so the terms in \reef{Expand} that result from expanding the tachyon poles do not appear in type 0 theory. In particular, there are no terms at second order in the Mandelstam variables. Hence, there is no four tachyon coupling at four-derivative order, and also no quartic term in the tachyon potential in type 0 theory. Hence, while the tachyon potential in bosonic string theory has a minimum, the tachyon potential in type 0 theory has no minimum.

\section{Discussion}

In this paper, we have shown that in the scheme where the non-derivative tachyon appears only in the tachyon potential, the higher-derivative classical effective action of bosonic string theory at order \(\alpha'^n\) can be separated into two sectors. One sector contains higher-derivative couplings of only massless fields at order \(\alpha'^n\). These couplings should be invariant under the restricted T-duality and consistent with S-matrix elements. This sector is called the massless sector. The other sector contains higher-derivative couplings of the tachyon and massless fields at order \(\alpha'^n\). These couplings should also be invariant under the restricted T-duality and consistent with S-matrix elements. This sector is called the tachyon sector.

These two sectors can be found by using a trick. We first constrain the minimal basis of massless and tachyon fields at order \(\alpha'^n\) to be invariant under Buscher rules supplemented by the most general higher-derivative corrections to all base-space massless fields. In this way, one can find the minimum number of T-duality-invariant multiplets. However, the minimal basis in which the tachyon potential in the equations of motion is ignored and the most general higher-derivative field redefinitions are used is not consistent with the S-matrix, because the tachyon is treated as a massless field; similarly, the most general corrections to the Buscher rules are not allowed, because they also treat the base-space tachyon as a massless field. We then impose further field redefinitions to convert the T-duality-invariant basis into new couplings that are invariant under the restricted T-duality transformations, in which the base-space metric and dilaton in the Buscher rules do not receive higher-derivative corrections. The restricted T-duality transformations would not treat the tachyon as a massless field; hence, the new couplings should be consistent with S-matrix elements. Such a field redefinition would not change the number of T-dual multiplets; it changes only the form of the couplings in each multiplet. This field redefinition can separate the T-duality-invariant couplings into the massless and tachyon sectors.

In order to find the massless sector, we use the following steps. Instead of using a field redefinition on the corresponding   T-dual multiplet in the minimal basis, we first find a basis of only massless fields. Such a basis can be found by using a restricted field redefinition that leaves the metric and dilaton unchanged. This restricted field redefinition does not treat the tachyon as a massless field and should therefore be consistent with S-matrix elements. We then fix its  coupling constants by constraining them to be invariant under the restricted T-duality transformations, in which there are no higher-derivative corrections to the base-space metric and dilaton. Using these steps, we have found the massless sector at order \(\alpha'^2\) in \reef{S2}, which contains 83 massless couplings. The tachyon sector has only two couplings: one involves four tachyon fields and the other involves six tachyon fields; they appear in \reef{TTTT}. We have shown that the massless sector is consistent with the expansion of the S-matrix element of two tachyons and two massless vertex operators, and found that the term in the tachyon sector involving four tachyons is reproduced by the expansion of the S-matrix element of four tachyon vertex operators. This fixes its coefficient as in \reef{4T} .

We have seen in the Introduction that the metric field redefinition \(\delta G^{\mu\nu}\) and the dilaton field redefinition \(\delta\Phi\) produce higher-derivative couplings that include, among other things, the non-derivative tachyon. This happens because the tachyon potential appears in the equations of motion of these fields. If one does not consider the tachyon potential in the equations of motion, then the tachyon is treated as a massless field. Therefore, the appropriate field redefinition should set \(\delta G^{\mu\nu} = 0\) and \(\delta\Phi = 0\). We have used such a restricted field redefinition to find the couplings in the massless sector. One may consider an arbitrary field redefinition \(\delta G^{\mu\nu}\) for the metric and a particular field redefinition for the dilaton such that \(\delta\Phi = \delta G^\mu{}_\mu/4\). In that case, the sum of the metric field redefinition and the dilaton field redefinition, \(\delta\Phi \Delta\Phi + \delta G^{\mu\nu}\Delta G_{\mu\nu}\), has no \(V(T)\). The corresponding corrections to the Buscher rules, i.e., \(\delta \bar{g}_{ab}\) and \(\delta \bar{\phi}\), may also be such that they satisfy the relation \(\delta\bar{\phi} = \delta \bar{g}^a{}_a/4\). In that case, one cannot separate the couplings into massless and tachyon sectors, because the corrections \(\delta \bar{g}_{ab}\) and \(\delta \bar{\phi}\) produce relations between terms that have only massless fields and terms that have the tachyon. Even though such a basis is consistent with the T-duality in which \(\delta\bar{\phi} = \delta \bar{g}^a{}_a/4\) at four-derivative order (see the four-derivative couplings in equation (12)  in \cite{Garousi:2026imu}), we have found that such a six-derivative basis is not consistent with T-duality in which \(\delta\bar{\phi} = \delta \bar{g}^a{}_a/4\) at six-derivative order.

The effective action of bosonic string theory for only massless fields has been found in \cite{Garousi:2019mca} to have 27 couplings with fixed coupling constants. Further field redefinitions convert these couplings into only 12 couplings \cite{Gholian:2023kjj}. In this paper, we have found that when both tachyon and massless fields are considered, the massless sector has 83 couplings with fixed coupling constants, which appear in \reef{S2}. Since this sector contains only the massless fields, one may use arbitrary field redefinitions to convert these 83 terms into the 12 terms that have been found in  \cite{Gholian:2023kjj}. However, such field redefinitions take us out of the scheme in which the non-derivative tachyon appears only in the tachyon potential. Now that we have the couplings in a particular scheme in which the non-derivative tachyon appears only in the tachyon potential, one may use arbitrary field redefinitions to convert the 83 massless couplings into 12 massless couplings. Such field redefinitions should also produce some derivative couplings in which the non-derivative tachyon appears. It would be interesting to find such a scheme in which the massless couplings are the 12 couplings that have been found in  \cite{Gholian:2023kjj}.

In finding the basis in which the non-derivative tachyon appears only in the tachyon potential, we have used the observation that there are no couplings with an odd number of tachyons in the effective action \cite{Garousi:2026nvz}. To clarify this point, consider the S-matrix element of four gravitons in both field theory and string theory. It has been shown in \cite{Metsaev:1986yb,Gholian:2023kjj} that the S-matrix element of four gravitons produced by covariant couplings in \reef{S0}, \reef{fourmax3}, and \reef{S2} is exactly reproduced by the momentum expansion of the following string theory amplitude \cite{Gross:1986iv,Kawai:1985xq}:
\beqa
A&\sim&\frac{\Gamma(-u/2)\Gamma(-s/2)\Gamma(-t/2)}{\Gamma(1+u/2)\Gamma(1+s/2)\Gamma(1+t/2)}K^B\bar{K}^B\,,\labell{Amp3}
\eeqa
where the open-string kinematic factor is
\beqa
K^B&=&\frac{st/4}{1+u/2}\z_1\cdot \z_3\z_2\cdot\z_4+\frac{su/4}{1+t/2}\z_2\cdot \z_3\z_1\cdot\z_4+\frac{ut/4}{1+s/2}\z_1\cdot \z_2\z_3\cdot\z_4+\cdots\,,\nn
\eeqa
where \(\zeta_i\) are the polarizations of the gauge bosons, and the dots represent terms in which the polarizations contract with momenta. There are similar relations for \(\bar{K}^B\). The graviton polarization is given by \(\epsilon_i = \zeta_i \bar{\zeta}_i\) for \(i = 1, 2, 3, 4\). The matching between the field theory amplitude and the string theory amplitude occurs when one expands the tachyon poles in \(K^B\) and \(\bar{K}^B\). Since the tachyon pole is expanded, the field theory cannot include couplings between two gravitons and one tachyon, even though the S-matrix element of two graviton and one tachyon vertex operators is nonzero. In fact, if the tachyon potential has a linear tachyon field, then all higher-derivative effective action terms can be removed by higher-derivative field redefinitions of the tachyon field. So it is mandatory to expand the tachyon pole, which results in an effective action with no linear tachyon field. This is exactly the same procedure that we have used in the expansion of the amplitude of two tachyons and two gravitons in \reef{AA}, and in the expansion of four tachyons in \reef{Expand}. This scenario is extended to any S-matrix element of tachyon and massless vertex operators, leading to the conclusion that the diffeomorphism-invariant field theory should not include an odd number of tachyons \cite{Garousi:2026nvz}.

It may be interesting to compare the expansion of S-matrix elements that we have used to produce the covariant effective action and its corresponding tachyon potential with the truncation method used in string field theory to produce the tachyon potential \cite{Belopolsky:1994sk,Yang:2005rx}. To find the quartic part of the tachyon potential, we consider expanding the tachyon pole in the amplitude of four tachyon vertex operators. This expansion produces a tower of higher-derivative couplings for four tachyons. Its zero-derivative part contributes to the tachyon potential. Moreover, the expansion of the massive poles in the amplitude produces higher-derivative couplings of four tachyons, starting from eight-derivative couplings. On the other hand, in string field theory, the couplings of two tachyons to all massless and massive fields are considered. Then, by using a truncation, one calculates the effect of massless and massive fields on the couplings of three tachyons or four tachyons \cite{Belopolsky:1994sk,Yang:2005rx}. In other words, in string field theory, one considers the effect of an infinite number of massive fields to find the effective tachyon potential, whereas in our expansion we consider only the effect of the tachyon pole on the tachyon potential. Even this tachyon pole has an infinite number of terms that affect only the derivative of the tachyon. Only one of them affects the tachyon potential.

We have shown that the expansion of the sphere-level S-matrix element of four tachyon vertex operators produces an expansion in terms of the old and new Mandelstam variables \reef{Expand}, which, up to three Mandelstam variables, is reproduced by higher-derivative couplings of four tachyons. The terms in the expansion that contain four Mandelstam variables should be reproduced by eight-derivative couplings. To find such couplings, one must consider the minimal basis at eight-derivative order and then impose invariance under the Buscher rules with arbitrary six-derivative corrections, the fixed two-derivative corrections \reef{d1}, and the four-derivative corrections that we have found in this paper (which, being lengthy expressions, we did not write explicitly). We leave this calculation for future work. However, the eight-derivative couplings in type 0 theory have no four- or six-derivative counterparts, so it is much easier to find them. The massless sector can be found by using an appropriate field redefinition on the corresponding NS-NS couplings of type II theory at eight-derivative order that have been found in \cite{Garousi:2020gio,Garousi:2020lof,Garousi:2022ghs}, such that the new massless couplings become invariant under the Buscher rules with restricted six-derivative corrections in which there are no corrections to the base-space metric and dilaton. The tachyon sector should be invariant under the Buscher rules. It is not difficult to see that the following couplings are invariant under the Buscher rules \reef{flac1}:
\beqa
\cL_T^{(3)}|_{\rm type0}&=&a\Big[\frac{1}{3}R_{\alpha }{}^{\epsilon }{}_{\beta 
}{}^{\varepsilon }R_{\gamma \epsilon \delta 
\varepsilon } \nabla^{\alpha }T \nabla^{\beta }T 
\nabla^{\gamma }T \nabla^{\delta }T + \frac{1}{6} 
\nabla^{\alpha }T \nabla^{\beta }T \nabla^{\gamma }T 
\nabla_{\delta }\nabla_{\gamma }R_{\alpha \beta } 
\nabla^{\delta }T \nn\\&&+ \nabla^{\beta }\nabla^{\alpha }T 
\nabla^{\gamma }\nabla_{\alpha }T \nabla_{\delta 
}\nabla_{\gamma }T \nabla^{\delta }\nabla_{\beta }T + 
\frac{4}{3}R_{\alpha \delta \beta \epsilon } 
\nabla^{\alpha }T \nabla^{\beta }T \nabla^{\delta 
}\nabla^{\gamma }T \nabla^{\epsilon }\nabla_{\gamma }T\nn\\&& -  
\frac{1}{3} \nabla^{\alpha }T \nabla^{\beta }T \nabla_{\gamma 
}R_{\alpha \delta \beta \epsilon } \nabla^{\gamma }T 
\nabla^{\epsilon }\nabla^{\delta }T\Big]+\cdots\,,
\eeqa
where $a$ is a constant and dots represents terms that have six tachyons and one term that has eight tachyons.  The four tachyon coupling produce the following result in the momentum space:
\beqa
A'^{(4)}_c&=&8ia\kappa^2\Big[(t'u')^2+(t's')^2+(u's')^2\Big]\,,
\eeqa
which is the same as the terms in expansion \reef{Expand} which has coeffient $\z(3)$. This fixes $a=\z(3)/128$. In the effective action of bosonic string theory there should be another T-dual multipelet which includes four-tachyon coupling as well as some other couplings. This terms should be invariant under Buscher rules that has two-derivative correstions \reef{d1} which relates  the last term in \reef{4T} to the eight-derivative couplings involving four tachyons. We leave the ditails evaluation of this term and other eight-derivative couplings for the future works.

The following closed form for the tachyon potential, which is consistent with \reef{VT1}, has been proposed in \cite{Garousi:2026imu}:
\[
V(T) = \frac{2}{\alpha'}(-1 + \cos T).
\]
If this is the correct form of tachyon potential, then the tachyon at the maximum, which occurs at \(T = 0\), condenses to the minimum of the potential, which occurs at \(T = \pi\). Around the minimum, the classical effective action admits an Anti–de Sitter solution with cosmological constant \(\Lambda = -4/\alpha'\). Using the AdS/CFT correspondence \cite{Maldacena:1997re,Gubser:1998bc,Witten:1998qj}, bosonic string theory around the tachyon minimum should then correspond to the IR limit of a conformal gauge theory on the boundary of the AdS space, and bosonic string theory around the maximum of the tachyon potential should correspond to the UV limit of the gauge theory. To provide justification for the above proposal for the tachyon potential, one requires the \(T^6\) term in the tachyon potential \reef{VT1}. The expansion of the sphere-level S-matrix element of six tachyon vertex operators should produce the six-tachyon couplings in \reef{4T}, as well as the coefficient of \(T^6\) in the tachyon potential. It would be interesting to find such a term and study the corresponding non-supersymmetric AdS/CFT correspondence.


\end{document}